\documentclass[aps,10pt,eqsecnum,preprint,nofootinbib,superscriptaddress]{revtex4}
\usepackage{amssymb,amsmath,amsthm,graphicx,amscd,mathtools}
\usepackage[mathscr]{eucal}
\usepackage{upgreek,enumerate,color,verbatim,comment,ulem,bigints,stackrel}
\usepackage{bm,pbox,cancel}
\usepackage[hidelinks]{hyperref}
\usepackage{orcidlink}

\begin{document}

\title{Non-Markovian Abraham-Lorentz-Dirac equation: Radiation Reaction without Pathology}  
\author{Jen-Tsung Hsiang\orcidlink{0000-0002-9801-208X}}
\email{cosmology@gmail.com}
\affiliation{Center for High Energy and High Field Physics, National Central University, Taoyuan 320317, Taiwan, ROC}
\author{Bei-Lok Hu\orcidlink{0000-0003-2489-9914}}
\email{blhu@umd.edu}
\affiliation{Maryland Center for Fundamental Physics and Joint Quantum Institute,  University of Maryland, College Park, Maryland 20742, USA}

\begin{abstract}
Motion of a point charge emitting radiation  in an electromagnetic field obeys the Abraham--Lorentz--Dirac (ALD) equation, with the effects of radiation reaction or self-force incorporated. This class of equations describing backreaction, including also the equations for gravitational self-force or Einstein's equation for cosmology driven by trace anomaly, contain  third-order derivative terms. They are known to have pathologies like the possession of runaway solutions, causality violation in pre-acceleration and the need for an extra second-order derivative initial condition. In our current program we reexamine this old problem from a different perspective, that of non-Markovian dynamics in open systems. This conceptual and technical framework has been applied earlier to the study of backreaction of quantum field effects on charge and mass motions and in early universe cosmology. Here we  consider a moving atom whose internal degrees of freedom, modeled by a harmonic oscillator, are coupled to a scalar field in the same manner as in scalar electrodynamics. Due to the way it is coupled to the charged particle, the field acts {effectively} like a supra-Ohmic environment, although {the field itself actually} has an Ohmic spectral density. We thus have cast the problem of radiation reaction to a study of the non-Markovian dynamics of a Brownian oscillator in a supra-Ohmic environment.  Our analysis shows that a) there is no need for specifying a second derivative for the initial condition; b) there is no pre-acceleration. These  undesirable features in conventional treatments arise from an inconsistent Markovian assumption: these equations were regarded as Markovian ab initio, not as a limit of the backreaction-imbued non-Markovian equation of motion. If one starts with the full non-Markovian dynamical equation and takes the proper Markovian limit judiciously, no harms are done. Finally, c) There is no causal relation between the higher-derivative term in the equation of motion and the existence of runaway solutions.  The runaway behavior is a consequence that the memory time in the environment is shorter than a critical value, which in the case of the charged particle, is the classical charge radius. If the charge  has an effective size greater than this critical value, its dynamics is stable. When this reasonable condition is met, radiation reaction understood and treated correctly in the non-Ohmic non-Markovian dynamics still obeys a third-order derivative equation, but it does not require a second derivative initial condition, and there is no pre-acceleration.
\end{abstract}
\maketitle

\baselineskip=18pt
\allowdisplaybreaks
\numberwithin{equation}{section}

\section{Introduction}

An accelerating electric charge emits  electromagnetic  radiation, a changing mass quadrupole emits gravitational radiation. Backreaction of this classical radiation on the moving charge or mass is known as  radiation reaction or self-force. The equation of motion for the moving charge including radiation reaction is the famous Abraham--Lorentz--Dirac (ALD) equation \cite{Abr,Lor,Dir,Rohrlich,GHW,BDR},  that for a mass including self-force is known as the Mino--Sasaki--Tanaka--Quinn--Wald (MST-QW) equation \cite{MST,QW}. Here we will focus on electromagnetic radiation reaction. Readers interested in gravitational self-force may find useful references in these representative papers and reviews \cite{Poisson,Gralla,Damour,Pound}.

As is well-known the ALD equation contains third-order derivatives with time, which requires the specification of three instead of two quantities - position, velocity and acceleration - as initial conditions. This opens the door to  fictitious force, pre-acceleration, and ensuing causality issues.  Numerous ingenious order reduction and iteration  schemes have been proposed to cope with finding physically meaningful and mathematically sound solutions.  Another well-known issue is the existence of runaway solutions. Attempts to tame them at late times  often lead to pre-acceleration problems at the initial time. These are referred to as the `pathologies' of radiation reaction. Over time, people have gotten used to living with them.

One reasonable  conceptual framework to explain (away) the origins of the {pathologies} and to find ways to lesson their harm is by invoking effective field theory (EFT) \cite{EFT,BurBk,Burgess,Donog,WeinbergEFTinf}. This is a vast field. For seminal papers and reviews on using EFT concepts or techniques for gravitational radiation and reaction problems, see, e.g., \cite{GolRot,GalTig,Porto,Levi}. One of the authors of this paper and collaborators have invoked EFT in previous work for some of the key issues raised here \cite{JH1,GH3}. In this work we propose a different way to deal with the problems with far-reaching implications, and to show that a proper treatment of the backreaction problem leads to a more satisfactory understanding of its underlying physics. The conceptual framework is quantum open systems (QOS)~\cite{CalHu97,Chou85,DeW86,Jor86,CH08,UW12,BrePet,RivHue,AK11,JR09} and the focus is on treating fully non-Markovian dynamics\footnote{There is no contradiction of QOS with effective field theory, since in our conceptual systematics, EFT is a subclass of open systems, where primary attention is paid to the system's energy scale and effective closure in the hierarchy of theories, but ignores the backreaction of  the often discarded sectors, that which do not affect the low energy theory's effectiveness. A small step towards placing EFT in the broader physical context and technical framework of QOS was taken in \cite{CalHu97} using the magnitude of noise representing the higher energy sectors, the ‘castaways’, as a measure of the effectiveness of a low energy EFT.}. We shall show that the pathologies arise from taking the Markovian limit improperly.

\subsection{Connection with work of similar veins}

Casting EFT in the QOS framework is a subject of importance in theoretical physics worthy of serious exploration.  Closer to what has been done, namely, using QOS concepts and nonequilibrium techniques (see, e.g., \cite{UW12,CH08}) to explore radiation phenomena, we mention two veins, one we are pursuing now, and the other, one of us has worked on with collaborators.

\paragraph{From quantum fluctuations to radiation and reaction}

In a series of recent papers starting with \cite{QRad}, we explored how fluctuations in a quantum field bring about quantum radiation from an atom, how its backreaction  engenders quantum dissipation in the {nonequilibrium} dynamics of the internal degrees of freedom of the atom. We also treated the same issues \cite{QRsq} for a squeezed state of the quantum field which enjoy applications to cosmological issues such as particle creation and structure formation. We then worked our way up to classical radiation and radiation reaction \cite{QRcoh} and showed the fallacy in the folklore of relating quantum fluctuations in a field with  classical radiation reaction. In this endeavor we invoked two main ingredients: one from quantum field theory to calculate the energy flux in a field, the other from quantum dissipative systems, namely, we demonstrated the existence of a {nonequilibrium} fluctuation--dissipation relation between quantum noise in the field and quantum radiation reaction. 

In this paper we turn our attention to unphysical behaviors in radiation reaction and show that properly seeking solutions to the full  non-Markovian dynamical equations of motion for the internal degree of freedom (idf) of the harmonic atom and correctly taking the Markovian limit can produce the equations of radiation reaction with no pathologies associated with the higher-derivative term. We can also identify the transition from stable to unstable solutions with the memory times of the bath.  More details are given at the end of this section and in the Summary section.

\paragraph{Derivation of radiation reaction equations via stochastic field theory}

Our present work shares the same root system with two earlier programs one of us developed with Johnson \cite{JH1}, Galley and collaborators \cite{GH1,GHL} based on the worldline influence functional formalism and stochastic field theory  in its use of QOS concepts and nonequilibrium quantum field theory techniques for the treatment of quantum fluctuations/noise, quantum radiation/dissipation, and classical radiation and radiation reaction/self-force. These earlier programs focused on the derivation of ALD--Langevin and MST--QW--Langevin equations. Here we analyze the solutions of the ALD--Langevin equation. The simple take home message is the same as what we said above, that one should start with solving the fully non-Markovian dynamical equations then take the proper Markovian limit with care rather than taking a shortcut to a Markovian equation and find the solutions carrying {pathologies}.

\paragraph{Non-Markovian dynamics}

We model the atom's idf by a quantum harmonic oscillator and use scalar electrodynamics as a model of the system-field interaction. We first show that a scalar electromagnetic field corresponds to a supra-Ohmic bath and derive the Langevin equation for the idf of the atom under the influence of the scalar field. We then focus on seeking the solutions to this equation taking into account the fully non-Markovian dynamics of the reduced system. 

Investigations of non-Markovian processes in QOS are still in their developmental stage. Began  three decades ago, notable landmarks include the derivation of a non-Markovian master equation for quantum Brownian motion (QBM)~\cite{HPZ92,HPZ93} (and its associated Fokker--Planck--Wigner equation~\cite{HalYu} and Langevin equation~\cite{CRV,CRV2}), experimental tests of the effects of  non-Markovian environments~\cite{Markus,Guo},  the many interesting proposals for the criteria and measures of non-Markovianity (see these reviews~\cite{nMrev,nMrev1,nMrev2,nMrev3}), and studies of non-Markovian behavior in various schemes of quantum information processing~\cite{LiuGoan,AnZhang,Wilson,EstPac}. For open system approach including non-Markovian dynamics considerations to radiation reaction problems, see \cite{FOC91,FLO85,FLO88}

Most of the papers addressing the non-Markovian criteria and measures are focused on the open system’s short time dynamics. To see more than just distinguishing the non-Markovian  features from the Markovian, one needs to go  beyond the initial transient stage.  One needs to follow the time evolution of the reduced system at  time ranges long enough to be able to see its fully non-Markovian dynamics~\cite{HAH22}. Serious efforts have been made to find solutions to the non-Markovian master or Langevin equations for quantum Brownian motion~\cite{CFdiss,Fleming,Fleming1,CRV,FordOC}.

\subsection{Pathologies in ALD Equations for Radiation Reaction} 

Motion of a charged particle in a field is a standard example of reduced dynamics in the open system framework, where the charged particle is the system of our interest and the electromagnetic field serves as its environment. Backreactions of the field enter into the equation of motion of the charged particle in the forms of radiation reaction (damping) and the Lorentz force. In a quantum field the Lorentz force is not only time-dependent and also fluctuating.

The electromagnetic field, due to the way it couples with the charged particle, acts like a supra-Ohmic environment, although the vector potential itself has an Ohmic spectral density, shown in Sec.~\ref{S:bgjvcher}. The frequency spectrum of the field typically ranges from 0 to infinity, but from physical considerations, a cutoff frequency exists which reflects the range of validity  of the electromagnetic theory or the structure of the system it interacts with. When the charge is assumed to be pointlike, the cutoff frequency is usually taken to infinity to simplify calculations, if  quantum phenomena  like pair creation of charged particles are excluded from one's considerations. The equation of motion for the point charge thus obtained is the ALD equation, which is known to contain  third-order time derivative of the position variable. It accounts for damping due to the back-reaction of emitted radiation in the field from the charge. The presence of this term leads to the following pathologies: 
\begin{enumerate}
	\item Third initial condition: since the equation of motion is a third-order ordinary differential equation, it needs three conditions to uniquely determine the solution, in contrast to two conditions in typical Newtonian dynamics.
	\item Runway behavior: the solution contains a component that grows unbounded exponentially.
	\item Pre-acceleration: to avoid the runaways, the ALD equation is often rephrased as an integro-differential equation. Although the corresponding solution does not exhibit the runaway behavior, the charge described by this solution will accelerate before the force is applied. It violates causality. 
\end{enumerate}
We show in this paper that the popular narrative about a point charge moving in an electromagnetic field is a special limiting case of the fully non-Markovian dynamics of the reduced system in the open systems  conceptualization and formulation, namely, the formal Markovian limit. From the broader perspectives of non-Markovianity in the open systems dynamics, most of these commonly encountered pathologies can be eliminated. Namely,  
\begin{enumerate}
	\item The presence of the third-order time derivative term is innocuous. There is no need for an additional third initial condition beyond the usual two.
	\item Presence of third-order time derivative term does not necessarily lead to runaways. The system's dynamical stability hinges more on the choice of the cutoff scale, whose inverse can also be understood as the memory time of the environment. The instability (runaway) emerges only when the memory time is shorter than a critical value.
	\item Pre-acceleration is not an issue in this context.
\end{enumerate}
To identify how the conventional formulation leads to these pathologies, we will in Sec.~\ref{S:bgjvcher} briefly go over the derivation of the equation of motion of a Brownian oscillator (Unruh-DeWitt detector) in electrodynamics in the open systems framework. In this way we can compare with the more proper and correct treatment from the fully non-Markovian dynamics, which is our central theme. 

In Sec.~\ref{S:fkgfdgfg}, we discuss the general characteristics and dynamical stability of the supra-Ohmic Brownian oscillator, coupled to a non-Markovian bath. We show the shared features in common with  a charged oscillator in scalar electrodynamics. Then in Sec.~\ref{S:tubgyidf} we use a specific examples of bath spectral density to illustrate the ideas discussed formally in Sec.~\ref{S:fkgfdgfg}. In Appendix~\ref{S:oetruihisurt} we briefly summarize the Ohmic dynamics for comparison, and in Appendix~\ref{S:rjubgdfg}, we address the ambiguities that arise  from treating the delta function in taking the Markovian limit. Then we provide more examples of bath spectral densities in Appendix~\ref{S:erbsdkf} for discussions in Sec.~\ref{S:fkgfdgfg}. In particular, we highlight some unusual properties of the hard-cutoff bath spectrum. In Appendix~\ref{S:kgbertdds} we comment on the higher-order supra-Ohmic systems.

\section{Brownian Oscillator in Scalar Electrodynamics}\label{S:bgjvcher}

{In this section we use the  textbook example of radiation from a charged oscillator in an electromagnetic field to illustrate the open systems approach in the derivation of the nonrelativistic ALD equation. We show that the reduced dynamics of a charged oscillator in an electromagnetic field,   by virtue of the velocity or derivative type of coupling, take on a form equivalent to that resulting from its interaction with a supra-Ohmic,  Markovian bath.}

\subsection{Nonrelativistic scalar ALD equation}
To avoid the complexity from the vectorial nature of the electromagnetic field, it is sufficient for our purpose to consider scalar electrodynamics. In the case we consider here, the results of both descriptions differ by a numerical factor, caused by the number of polarizations, angular weight, and the choice of the unit system in electromagnetism. The scalar field discussed here is analogous to the vector potential in electromagnetism, and that the equation of motion of the scalar charge oscillator is parallel to the Lorentz equation, including  the consideration of the reactive force from (scalar field) radiation.

Suppose that the internal degree of freedom $\chi(t)$ of an Unruh-DeWitt detector is modeled by a one-dimensional, nonrelativistic oscillator, which carries the scalar charge $e$. The external or mechanical (edf)  degree of freedom $\bm{z}(t)$ of the detector may follow a given trajectory in $3+1$ Minkowski space, but for our purpose with focus on the internal degrees of freedom   we can assume it will stay at rest for simplicity. The internal degree of freedom (idf) of the detector is coupled to an ambient  massless scalar field $\phi(\bm{x},t)$. The oscillator's velocity $\dot{\chi}(t)$ is coupled to the field $\phi(\bm{x},t)$ in the same way as in electrodynamics. For the moment assume that the detector is pointlike, and the spectrum of the field extends to infinity. We shall return to this assumption later.

The action of  the oscillator-field system as described is given by 
\begin{equation}\label{E:ksjgbdkf}
	S=\int\!ds\;\biggl[\frac{m_{\textsc{b}}^{\vphantom{2}}}{2}\dot{\chi}^{2}(s)-\frac{m_{\textsc{b}}^{\vphantom{2}}\omega_{\textsc{b}}^{2}}{2}\,\chi^{2}(s)\biggr]+e\int\!d^{4}y\;\phi(\bm{y},s)\,\dot{\chi}(s)\,\delta(\bm{y}-\bm{z})+\int\!d^{4}y\;\Bigl\{\frac{1}{2}\bigl[\partial_{s}\phi(\bm{y},s)\bigr]^{2}-\frac{1}{2}\bigl[\partial_{\bm{y}}\phi(\bm{y},s)\bigr]^{2}\Bigr\}
\end{equation}
where $y^{\mu}=(s,\bm{y})$ is the external (spacetime) coordinate. The overhead dot represents taking the derivatives with respect to time. The mass and the oscillator's natural frequency take on bare values, denoted by the subscript $\textsc{b}$, for the moment. Divergent contributions from the oscillator-field interaction will be dealt with accordingly. The action \eqref{E:ksjgbdkf} is seen to describe essentially the scalar version of electromagnetic interaction, with $\phi$ playing the role of the electromagnetic vector potential, and $e\dot{\chi}$ the role of (internal) electric current.

This action produces a simultaneous set of equations of motion
\begin{align}
	m_{\textsc{b}}^{\vphantom{2}}\,\ddot{\chi}(t)+m_{\textsc{b}}^{\vphantom{2}}\omega_{\textsc{b}}^{2}\,\chi(t)&=-e\,\dot{\phi}(\bm{z},t)\,,\label{E:rkgbfsd1}\\
	\ddot{\phi}(\bm{x},t)-\partial^{2}_{\bm{x}}\phi(\bm{x},t)&=e\,\dot{\chi}(t)\,\delta(\bm{x}-\bm{z})\,,\label{E:rkgbfsd2}
\end{align}
in analogy with the Lorentz equation of the charge and the wave equation of the vector potential, respectively. If the oscillator-field interaction is switched on at $t=0$, then the formal solution to \eqref{E:rkgbfsd2} is given by
\begin{align}\label{E:rygivfdgdd}
	\phi(\bm{x},t)=\phi_{h}(\bm{x},t)+e\int_{0}^{t}\!d^{4}y\;G_{R,0}^{(\phi)}(\bm{x},t;\bm{y},s)\,\dot{\chi}(s)\,\delta(\bm{y}-\bm{z})\,,
\end{align}
where $\phi_{h}(\bm{x},t)$ denotes the homogeneous solution to the wave equation, \eqref{E:rkgbfsd2}. The subscript $0$ in $G_{R,0}^{(\phi)}(\bm{x},t;\bm{y},s)$ tells us that it is the  retarded Green's function of the free scalar field $\phi_{h}$, not the full field $\phi$. The second term on the righthand side of \eqref{E:rygivfdgdd} gives the radiation field emitted by the charge oscillator. We then substituted \eqref{E:rygivfdgdd} into \eqref{E:rkgbfsd1} to find an equation of motion for the charged oscillator under the influence of the environment,
\begin{align}
	m_{\textsc{b}}^{\vphantom{2}}\,\ddot{\chi}(t)+m_{\textsc{b}}^{\vphantom{2}}\omega_{\textsc{b}}^{2}\,\chi(t)+e^{2}\frac{\partial}{\partial t}\int_{0}^{t}\!ds\;G_{R,0}^{(\phi)}(\bm{z},t;\bm{z},s)\,\dot{\chi}(s)=-e\,\frac{\partial}{\partial t}\hat{\phi}_{h}(\bm{z},t)\,.\label{Ejhgdf}
\end{align}
The term on the righthand side plays the role of the Lorentz force, while the nonlocal expression will account for radiation damping {by the scalar field radiation}. In the coincident spatial limit, we usually write $G_{R,0}^{(\phi)}(\bm{z},t;\bm{z},s)$ as $G_{R,0}^{(\phi)}(t-s)$ to avoid cluttering of notations.

Although the  {scalar} field itself has an Ohmic spectrum, the velocity coupling can induce supra-Ohmic effects by the presence of additional time derivatives in the nonlocal term in \eqref{Ejhgdf}. This can be made more explicit if we write the nonlocal term as
\begin{align}
	\frac{\partial}{\partial t}\int_{0}^{t}\!ds\;G_{R,0}^{(\phi)}(t-s)\,\dot{\chi}(s)&=G_{R,0}^{(\phi)}(0)\,\dot{\chi}(t)+\int_{0}^{t}\!ds\;\partial_{t}G_{R,0}^{(\phi)}(t-s)\,\dot{\chi}(s)\notag\\
	&=-\ddot{\Gamma}^{(\textsc{e})}(0)\,\chi(t)+\ddot{\Gamma}^{(\textsc{e})}(t)\,\chi(0)+\int_{0}^{t}\!ds\;\partial^{3}_{s}\Gamma^{(\textsc{e})}(t-s)\,\chi(s)\,.
\end{align}
To place it in the context of {supra-Ohmic} dynamics we have used the substitution $G_{R,0}^{(\phi)}(\tau)=-\partial_{\tau}\Gamma^{(\textsc{e})}(\tau)$ where $\Gamma^{(\textsc{e})}(\tau)$ is an Ohmic kernel function, discussed in fuller details in Appendix~\ref{S:oetruihisurt}. The third-order time derivative in the last term will give a factor of $\kappa^{3}$ in the Fourier representation of $\Gamma^{(\textsc{e})}(\tau)$,
\begin{equation}
	\partial^{3}_{\tau}\Gamma^{(\textsc{e})}(\tau)=\partial^{3}_{\tau}\int_{-\infty}^{\infty}\!\frac{d\kappa}{2\pi}\;\frac{I(\kappa)}{2\pi\kappa}\,e^{-i\kappa\tau}=i\int_{-\infty}^{\infty}\!\frac{d\kappa}{2\pi}\;\frac{\kappa^{3}I(\kappa)}{2\pi\kappa}\,e^{-i\kappa\tau}\,,
\end{equation}
and thus $\partial^{3}_{\tau}\Gamma^{(\textsc{e})}(\tau)$ accounts for the supra-Ohmic effect. The function $I(\kappa)$ is the spectral density of the scalar field $\phi$ and has the Ohmic form.

In the Markovian limit where the cutoff frequency $\Lambda$ is taken to infinity, {the kernel function $\Gamma^{(\textsc{e})}(\tau)$ is given by 
\begin{equation}
    \Gamma^{(\textsc{e})}(\tau)=\frac{1}{2\pi}\,\delta(\tau)\,,
\end{equation}   
and thus} the retarded Green's function of the free field $G_{R,0}^{(\phi)}(\tau)$ {in the spatial coincident limit} has a very simple form
\begin{equation}
	G_{R,0}^{(\phi)}(\tau)=-\theta(\tau)\,\frac{1}{2\pi}\,\dot{\delta}(\tau)\,,\label{E:trigdfse}
\end{equation}
where $\theta(\tau)$ is the Heaviside function. This is highly singular. Hereafter we will discuss the confusions and difficulty arisen in interpreting the supra-Ohmic dynamics, if the Markovian limit of this kernel function is treated prima facie as a Dirac-delta function.

Before formally finding the solution,  {let us first} address the renormalization of the parameters. Inserting \eqref{E:trigdfse} into the non-local term yields
\begin{align}
	e^{2}\frac{\partial}{\partial t}\int_{0}^{t}\!ds\;G_{R}^{(\phi)}(t-s)\,\dot{\chi}(s)&=\frac{e^{2}}{2\pi}\frac{\partial}{\partial t}\int_{0}^{t}\!ds\;\Bigl[\frac{\partial}{\partial s}\delta(t-s)\Bigr]\,\dot{\chi}(s)\notag\\
	&=\frac{e^{2}}{2\pi}\frac{\partial}{\partial t}\biggl[\delta(0)\,\dot{\chi}(t)-\delta(t)\,\dot{\chi}(0)-\int_{0}^{t}\!ds\;\delta(t-s)\,\ddot{\chi}(s)\biggr]\notag\\
	&=\frac{e^{2}}{2\pi}\biggl[\delta(0)\,\ddot{\chi}(t)-\dot{\delta}(t)\,\dot{\chi}(0)-\partial_{t}\int_{0}^{t}\!ds\;\delta(t-s)\,\ddot{\chi}(s)\biggr]\notag\\
	&=\frac{e^{2}}{2\pi}\biggl[\delta(0)\,\ddot{\chi}(t)-\dot{\delta}(t)\,\dot{\chi}(0)-\delta(t)\,\ddot{\chi}(0)-\frac{1}{2}\dddot{\chi}(t)\biggr]\,.\label{E:jguywzd}
\end{align}
Here we have given more details in derivations to show the subtleties: (A) Probably better accepted, the integral $\displaystyle\int_{0}^{t}\!ds\;\delta(t-s)\,\ddot{\chi}(s)$ only gives half of the contribution because heuristically half of the ``peak'' of the delta function is used;  however, (B) it is less known that when a delta function is involved in an expression like $\displaystyle\partial_{t}\int_{0}^{t}\!ds\;\delta(t-s)\,\ddot{\chi}(s)$ should give a result like
\begin{equation}
	\partial_{t}\int_{0}^{t}\!ds\;\delta(t-s)\,\ddot{\chi}(s)=\delta(t)\,\ddot{\chi}(0)+\frac{1}{2}\dddot{\chi}(t)\,.
\end{equation}
The first term on the righthand side may be argued to be irrelevant when $t>0$, but {its very existence} is needed for a consistent (formal) Markovian limit and in the consideration of the number of the initial conditions. It is often omitted inadvertently. A more detailed discussion on consistent implementation and interpretation of the delta function in the context of taking the Markovian limit is discussed in Appendix~\ref{S:rjubgdfg}. A short conclusion is that we should interpret the delta function by its asymptotic form, commensurate with the specified physical setting.

The first term in \eqref{E:jguywzd} corresponds to mass renormalization, so by the following identifications
\begin{align}
	m_{\textsc{b}}^{\vphantom{2}}&\mapsto m_{\textsc{p}}^{\vphantom{2}}=m_{\textsc{b}}^{\vphantom{2}}+\frac{e^{2}}{2\pi}\,\delta(0)\,,&m_{\textsc{b}}^{\vphantom{2}}\omega_{\textsc{b}}^{2}&\mapsto m_{\textsc{p}}^{\vphantom{2}}\omega_{\textsc{p}}^{2}\,,
\end{align}
the equation of motion \eqref{Ejhgdf} then reduces to\footnote{One may construct an equation of motion which has the form $m_{\textsc{p}}^{\vphantom{2}}\,\ddot{\chi}(t)+m_{\textsc{p}}^{\vphantom{2}}\omega_{\textsc{p}}^{2}\,\chi(t)-2m_{\textsc{p}}^{\vphantom{2}}\gamma\,\dddot{\chi}(t)=-e\,\frac{\partial}{\partial t}\hat{\phi}_{h}(\bm{z},t)$ even when $t=0$, but it describes slightly different dynamics from \eqref{E:irutyvs} and will not correspond to the formal Markovian limit of \eqref{E:rtgufyvdhf} or \eqref{E:zrubgdh}. Instead, we will have the Laplace transform of the solution given by
\begin{align*}
	\tilde{\chi}(\sigma)&=\frac{(1-2\gamma\sigma)\sigma}{\sigma^{2}+\omega_{\textsc{p}}^{2}-2\gamma\,\sigma^{3}}\,\chi(0)+\frac{1-2\gamma\sigma}{\sigma^{2}+\omega_{\textsc{p}}^{2}-2\gamma\,\sigma^{3}}\,\dot{\chi}(0)-\frac{2\gamma}{\sigma^{2}+\omega_{\textsc{p}}^{2}-2\gamma\,\sigma^{3}}\,\ddot{\chi}(0)+\text{the particular solution}\,,
\end{align*}
slightly different from \eqref{E:kghbfg}.}
\begin{align}
	m_{\textsc{p}}^{\vphantom{2}}\,\ddot{\chi}(t)+m_{\textsc{p}}^{\vphantom{2}}\omega_{\textsc{p}}^{2}\,\chi(t)-2m_{\textsc{p}}^{\vphantom{2}}\gamma\,\dddot{\chi}(t)-4m_{\textsc{p}}^{\vphantom{2}}\gamma\Bigl[\dot{\delta}(t)\,\dot{\chi}(0)+\delta(t)\,\ddot{\chi}(0)\Bigr]=-e\,\frac{\partial}{\partial t}\hat{\phi}_{h}(\bm{z},t)\,,\label{E:irutyvs}
\end{align}
{with $\gamma=e^{2}/(8\pi m)$.} Here we see the appearance of $-\dddot{\chi}(t)$. It accounts for radiation damping taking the form of a self-force. It implies that the solution may depend on $\ddot{\chi}(0)$, in addition to $\chi(0)$ and $\dot{\chi}(0)$. Furthermore, this equation of motion itself contains a few terms that explicitly depends on the initial conditions. Their presence naturally arises in the framework of open systems {when we re-write \eqref{E:rygivfdgdd} as \eqref{E:irutyvs} in terms of physical parameters by means of} the repeated integrations by parts in \eqref{E:jguywzd}. Emphatically they are needed in correctly treating the initial-condition issues. When $t>0$, these terms disappear, and we arrive at the nonrelativistic ALD equation. The appearance of the $\dddot{\chi}(t)$ term in the ALD equation which is not seen in the typical Newtonian dynamics has drawn much attention to the invention of schemes to  revise the equation of motion by,  order reduction, interpretation of the solutions, and discussions of their consequences.

\subsection{Initial conditions: specification of $\ddot{\chi}(0)$ unnecessary in ALD equation}
The formal solution to \eqref{E:irutyvs} can be given by means of the Laplace transformation. However, there is a catch. The Laplace transformation of the delta function is typically assigned to be 1 because the lower bound in \eqref{E:gbuess} is replaced by $0^{-}$ in the standard reference. If we use this convention, we find
\begin{align}\label{E:kghbfg}
	\tilde{\chi}(\sigma)&=\frac{(1-2\gamma\sigma)\sigma}{\sigma^{2}+\omega_{\textsc{p}}^{2}-2\gamma\,\sigma^{3}}\,\chi(0)+\frac{1+2\gamma\sigma}{\sigma^{2}+\omega_{\textsc{p}}^{2}-2\gamma\,\sigma^{3}}\,\dot{\chi}(0)+\frac{2\gamma}{\sigma^{2}+\omega_{\textsc{p}}^{2}-2\gamma\,\sigma^{3}}\,\ddot{\chi}(0)\notag\\
	&\qquad\qquad\qquad\qquad\qquad\qquad\qquad\qquad\qquad\qquad\qquad\qquad+\text{the particular solution}\,,
\end{align}
where the particular solution denotes contribution from the force term, and the Laplace transform $\tilde{f}(\sigma)$ of a function $f(t)$ is defined by
\begin{equation}\label{E:gbuess}
	\tilde{f}(\sigma)=\int_{0}^{\infty}\!dt\;e^{-t\sigma}\,f(t)\,,
\end{equation}
with $\operatorname{Re}\sigma>0$, and the inverse transform is given by 
\begin{equation}
	f(t)=\frac{1}{2\pi i}\int_{C}\!d\sigma\;\tilde{f}(\sigma)\,e^{t\sigma}\,,
\end{equation}
where the closed contour $C$ encompasses all poles of $\tilde{f}(\sigma)$. {Sometimes, we use the notation like $\pounds f(t)$ to denote applying the Laplace transformation to the function $f(t)$.} We see that the solution in \eqref{E:kghbfg} needs the input from three initial conditions $\chi(0)$, $\dot{\chi}(0)$ and $\ddot{\chi}(0)$.

However, from the discussion in Appendix~\ref{S:rjubgdfg}, the following seems to us more consistent with the Markovian limit, if we still set the lower bound in \eqref{E:gbuess} to 0, and heuristically imagine that only half of the delta function ``peak'' is used to evaluate the Laplace transformation. Then the third subtlety in this supra-Ohmic Markovian dynamics is that (C) we will find the Laplace transform of the delta function is $1/2$, and likewise the Laplace transform of $\dot{\delta}(t)$ is $-\delta(0)+\dfrac{\sigma}{2}$, instead of $\sigma$. Using this algorithm, we obtain instead that
\begin{align}\label{E:kgjdbf}
	\tilde{\chi}(\sigma)=\frac{(1-2\gamma\sigma)\sigma}{\sigma^{2}+\omega_{\textsc{p}}^{2}-2\gamma\,\sigma^{3}}\,\chi(0)+\frac{1-4\gamma\sigma\,\delta(0)}{\sigma^{2}+\omega_{\textsc{p}}^{2}-2\gamma\,\sigma^{3}}\,\dot{\chi}(0)+\text{the particular solution}\,.
\end{align}
This result depends on only two initial condition and conforms to Newtonian dynamics. The appearance of $\delta(0)$ may look alarming, but its presence is necessary when we take the Markovian limit of non-Markovian dynamics, as will be shown later. Moreover, this solution is purely formal and is the runaway kind. From the fuller perspective of the non-Markovian dynamics we shall present below, the Markovian limit of this supra-Ohmic non-Markovian equation of motion is inherently unstable and it is not surprising that it possesses   runaway solutions. Finally, as we stress earlier that it will be more physically consistent to interpret the delta function by its asymptotic form. Here we remind again that the runaway behavior of the solution has nothing to do with the presence of $\delta(0)$ or its asymptotic form; the runaway results from the existence of the positive real root in $\sigma^{2}+\omega_{\textsc{p}}^{2}-2\gamma\,\sigma^{3}=0$.

Earlier in \eqref{E:jguywzd}, we mentioned that a term $-\dfrac{e^{2}}{2\pi}\delta(t)\,\ddot{\chi}(0)$ is often omitted. The restoration of this term is important because without its presence, the solution will still depend on the initial condition $\ddot{\chi}(0)$ no matter what convention of the Laplace transformation of the delta function we take.

The previous discussions on treating the delta function expressions, in particular in terms of its asymptotic form, will be better motivated and justified once we learn more about supra-Ohmic, non-Markovian dynamics in Sec.~\eqref{S:fkgfdgfg}.

\subsection{Runaway behavior mitigated by memories in the {non-Markovian bath} field}
The salient features revealed in our analysis of the dynamics of the charge particle in scalar electrodynamics are:  1) the appearance of the $\dddot{\chi}(t)$ in the equation motion does not always imply dynamical instability and the necessity of the additional initial condition $\ddot{\chi}(0)$ to uniquely determine the solution, 2) instability arises corresponding to the length of the memory time of the bath, and 3) the description of the Markovian limit in the supra-Ohmic dynamics is more complicated than in the Ohmic case. The {meaning of the} Markovian limit can only be reached and better understood by adopting a broader scope, starting from the full non-Markovian dynamics of the reduced system, as we shall show in the next section.

The non-Markovianity, or the finite cutoff scale, in the spectrum of the field can naturally emerge from various considerations. For example, it can be induced by the finite-size effect of the charge particle, or by the scale inherited in the modeling of the field configuration, such as in the presence of a dielectric or under spatial confinement. It may even reflect the energy scale associated with the validity of the theory. We will see that in the current discussion $2\gamma=e^{2}/(4\pi m_{\textsc{p}})$ plays  such a role. It gives the ``radius'' of the charge and defines a critical scale so that when the length scale corresponding to the memory time  is shorter than this radius, the dynamics of the charge begins to show runaway behavior. This is yet another reason why one should  refrain from treating the charge as a point particle.

This observation, that the charge should be treated as an extended object, is not new~\cite{Yaghjian,Erber,MonShar}. We reached this conclusion from the memory time or non-Markovianity considerations. When the size of the charge is larger than the classical radius there are no runaway solutions or pre-acceleration issues. In fact, these difficulties can be effectively mitigated by introducing a suitable scale into the theory or the calculations such as, for example,  the switching-on time scale of the charge-field interaction, the length scale in the electron form factor due to the shape or the charge distribution, or the finite-difference numerical algorithm\footnote{This is perhaps a good point to compare the work of Ref.~\cite{FOC91} with the present treatment:  Both start the non-Markovianity consideration  from  an equation of motion like \eqref{Ejhgdf}, but then the two approaches differ. The authors of~\cite{FOC91} convert the integro-differential equation \eqref{Ejhgdf} into an ordinary, but higher-order differential equation to make connection with the ALD equation. This is possible for the specific form of spectral density chosen there. However, for more general bath spectral functions,  it is not a simple task to make such a conversion. These authors identify the role of the cutoff, the origin of non-Markovianity, in dynamical stability based on causality requirement. In contrast, we stay in the non-Markovian formalism, and this allows us to address instability in a more general context in terms of the factor $1-8\pi\gamma\,\Gamma^{(\textsc{e})}(0)$, discussed in Sec.~\ref{S:fkgfdgfg}, and to better identify subtleties surrounding the Markovian limit in supra-Ohmic dynamics.}~\cite{JH1,FOC91,Pena,GLR,Lan}.

In the next section, we shall focus on a class of non-Markovian dynamics   often encountered in Brownian motion, and provide a systematic elaboration on its dynamical characteristics. In addition, we will establish its connection with the (nonrelativistic) non-Markovian ALD equation and then discuss its Markovian limit. We also provide a few examples to give concrete pictures of these dynamical features.

\section{Non-Markovian dynamics in supra-Ohmic baths}\label{S:fkgfdgfg} 
{We start with a general discussion of the non-Markovian dynamics in a supra-Ohmic bath, and identify important but often neglected features in the nonequilibrium dynamics of the supra-Ohmic non-Markovian system in the context of decoherence and entanglement dynamics, such as, dynamical stability, and dependence on the initial conditions. Later in Sec.~\ref{S:jhvgdjdfg} we will discuss its connection with the issue of radiation reaction of a moving charged particle in an electromagnetic field invoked in the previous section. In addition, for reader's convenience, in Appendix~\ref{S:oetruihisurt} we summarize the non-Markovian dynamics in an Ohmic bath for comparison.}

\subsection{Supra-Ohmic non-Markovian Langevin equation}
An intuitive and likely the simplest way to implement non-Markovian dynamics in a supra-Ohmic bath  is to choose a standard class of equation of motion for Brownian motion
\begin{equation}\label{E:zrubgdh}
	m_{\textsc{b}}^{\vphantom{2}}\,\ddot{\chi}(t)+m_{\textsc{b}}^{\vphantom{2}}\omega_{\textsc{b}}^{2}\,\chi(t)-e^{2}\int_{0}^{t}\!ds\;G_{R,0}^{(\textsc{e})}(t-s)\,\chi(s)=e\,\xi(t)\,,
\end{equation}
but to implement the spectral density of the bath by a supra-Ohmic form   $J(\kappa)=\kappa^{\lambda}\,\mathcal{P}_{\Lambda}(\kappa)$ with $\lambda>1$. Eq~\eqref{E:zrubgdh} describes the motion of a Brownian oscillator, which is the reduced system of interest to us, coupled to a supra-Ohmic thermal Gaussian bath. The influences of the bath are summarized into a noise force $\xi(t)$ and a nonlocal expression in \eqref{E:zrubgdh} that depends on the retarded Green's function of the free bath. The latter contains the reaction to the disturbance to the bath generated by the oscillator when it is driven by the noise force, and also includes corrections to or renormalization of the oscillator's physical parameters. The retarded Green's function of the free bath is given by
\begin{equation}
	G_{R,0}^{(\textsc{e})}(\tau)=i\,\theta(\tau)\int_{0}^{\infty}\!\frac{d\kappa}{2\pi}\;\frac{J(\kappa)}{2\pi}\,\Bigl[e^{-i\kappa\tau}-e^{+i\kappa\tau}\Bigr]\,,
\end{equation}
where $\theta(\tau)$ is the Heaviside theta function. To comply with the fluctuation-dissipation relation of the free bath, the noise $\xi(t)$ of the bath is required to satisfy the Gaussian statistics:
\begin{align}\label{E:dgbdfkgdf}
	\langle\xi(t)\rangle&=0\,,&&\text{and}&\frac{1}{2}\langle\bigl\{\xi(t),\xi(t')\bigr\}\rangle&=G_{H,0}^{(\textsc{e})}(\tau)=\int_{-\infty}^{\infty}\!\frac{d\kappa}{2\pi}\;\frac{J(\kappa)}{4\pi}\coth\frac{\beta\kappa}{2}\,e^{-i\kappa\tau}\,,
\end{align}
where $\tau=t-t'$, and $\beta$ is the inverse initial bath temperature. Here we will only deal with the $\lambda=3$ case. We will give a brief description of the higher supra-Ohmic system in Appendix~\ref{S:kgbertdds}.

Since from time to time, we would like to compare the supra-Ohmic dynamics with the Ohmic one, we will express the supra-Ohmic kernel functions in terms of $\Gamma^{(\textsc{e})}(\tau)$, which is often seen in the context of Ohmic Brownian motion. The kernel function $\Gamma^{(\textsc{e})}(\tau)$ takes the form
\begin{equation}\label{E:wwww}
	\Gamma^{(\textsc{e})}(\tau)=\int_{-\infty}^{\infty}\!\frac{d\kappa}{2\pi}\;\frac{I(\kappa)}{2\pi\kappa}\,e^{-i\kappa\tau}\,,
\end{equation}
such that supra-Ohmic spectral density $J(\kappa)$ is related to the Ohmic spectral density $I(\kappa)$ by $J(\kappa)=\kappa^{2}\,I(\kappa)$, and the supra-Ohmic retarded Green's function $G_{R,0}^{(\textsc{e})}(\tau)$ is expressed as
\begin{equation}\label{E:tgusyirth}
	G_{R,0}^{(\textsc{e})}(\tau)=\theta(\tau)\,\frac{\partial^{3}}{\partial\tau^{3}}\Gamma^{(\textsc{e})}(\tau)\,.
\end{equation}
For a bath consisting of a continuum of modes, the kernel functions can be potentially ill-defined. To circumvent this ambiguity, the function $\mathcal{P}_{\Lambda}(\kappa)$ is introduced to do the job. In this paper, we assume that it contains only one scale, the cutoff frequency $\Lambda$. The contributions from the modes of the bath whose frequencies are higher than $\Lambda$ will be suppressed. We also assume that $\mathcal{P}_{\Lambda}(\kappa)$  is an even function of $\kappa$, and approaches unity when $\Lambda$ goes to infinity. {This is necessary to ensure a unique Markovian limit, independent of the functional forms of the spectral densities}. The following three common forms of $\mathcal{P}_{\Lambda}(\kappa)$ will be discussed in this paper,
\begin{align}
	\mathcal{P}_{\Lambda}(\kappa)&=e^{-\frac{\lvert\kappa\rvert}{\Lambda}}\,,&\mathcal{P}_{\Lambda}(\kappa)&=\frac{\Lambda^{2}}{\Lambda^{2}+\kappa^{2}}\,,&\mathcal{P}_{\Lambda}(\kappa)&=\Theta(\Lambda-\kappa)\,\Theta(\kappa+\Lambda)\,.
\end{align}
In the first case, the high frequency modes are exponentially suppressed, while the second form, called Lorentzian, only algebraically. Thus in certain cases, the latter may not be enough to ease off the divergence due to the high-frequency bath modes. The last case, called hard cutoff, is probably the most na\"ive way to apply the cutoff. Any mode with a frequency higher than $\Lambda$ is outright discarded. We mention this case because we want to warn the danger of it having some peculiar, undesirable properties {when $\Lambda$ is not extremely large.} {The Lorentz spectrum will be treated in Sec.~\ref{S:tubgyidf}, and the other two examples are relegated to Appendix~\ref{S:erbsdkf}.}

Due to the presence of finite $\Lambda$ in the spectral density, the nonlocal expression in \eqref{E:zrubgdh} cannot be reduced to a local form\footnote{For example, if the kernel function $\Gamma^{(\textsc{e})}(\tau)$ is proportional to $\delta(\tau)$, then the integral expression like $\displaystyle\int_{0}^{t}\!ds\;\delta(t-s)\,\chi(s)$ reduces to $\dfrac{1}{2}\chi(t)$, local in term. We do not mean to re-write the whole {nonlocal} equation of motion, into the so-called time-local form. The latter can in principle be achieved for the linear non-Markovian equation of motion.}, and thus it will allow the past history to affect the current state of motion. Then for any finite $\Lambda$, the dynamics described by \eqref{E:zrubgdh} will be history-dependent, and hence is non-Markovian\footnote{{In this paper, non-Markovianity has a broader connotation, referring to memories in the open system's dynamics, not an approximation used in expressing the two-point function. For example, the Hadamard function $G_{H,0}^{(\textsc{e})}(\tau)$ in \eqref{E:dgbdfkgdf} in the high-temperature limit is approximately given by (the derivative of) the delta function, and is said to have a Markovian form. We refrain from this  narrower sense of usage}.}. From the duality properties of the Fourier transformation, we will expect that $\Lambda^{-1}$ plays the role of memory time. It determines how far back past history of the motion can affect the current motion. Taking the Markovian limit by letting $\Lambda\to\infty$ can then be understood as letting the memory time approach zero, and the system dynamics becomes memoryless; it forgets everything that happens before the current moment.

We first take care of the renormalization issue in the supra-Ohmic equation of motion \eqref{E:zrubgdh}. The supra-Ohmic spectral density of the bath implies that the non-local term in \eqref{E:zrubgdh} will have a more serious {divergence} issue than the Ohmic bath. Inserting \eqref{E:tgusyirth} into \eqref{E:zrubgdh}, we can write the nonlocal term as
\begin{align}
	&\quad-e^{2}\int_{0}^{t}\!ds\;G_{R}^{(\textsc{e})}(t-s)\,\chi(s)\notag\\
	&=e^{2}\int_{0}^{t}\!ds\;\frac{\partial^{3}}{\partial s^{3}}\Gamma^{(\textsc{e})}(t-s)\,\chi(s)\label{E:ruthit}\\
	&=e^{2}\biggl\{\ddot{\Gamma}^{(\textsc{e})}(0)\,\chi(t)-\ddot{\Gamma}^{(\textsc{e})}(t)\,\chi(0)+\dot{\Gamma}^{(\textsc{e})}(0)\,\dot{\chi}(t)-\dot{\Gamma}^{(\textsc{e})}(t)\,\dot{\chi}(0)+\Gamma^{(\textsc{e})}(0)\,\ddot{\chi}(t)-\Gamma^{(\textsc{e})}(t)\,\ddot{\chi}(0)\biggr\}\notag\\
	&\qquad\qquad\qquad\qquad\qquad-e^{2}\int_{0}^{t}\!ds\;\Gamma^{(\textsc{e})}(t-s)\,\dddot{\chi}(s)\,.\label{E:dkhgfssg}
\end{align}
The term proportional to $\ddot{\chi}(t)$ in \eqref{E:dkhgfssg} will be absorbed into mass renormalization. This is not seen in the treatment of typical Ohmic Brownian motion, where only the natural frequency needs renormalizing. The term proportional to $\chi(t)$ is then associated with the frequency renormalization in the supra-Ohmic case, such that the physical mass and the physical frequency are respectively given by
\begin{align}
	m_{\textsc{p}}&=m_{\textsc{b}}+e^{2}\Gamma^{(\textsc{e})}(0)\,,&m^{\vphantom{2}}_{\textsc{p}}\omega^{2}_{\textsc{p}}&=m^{\vphantom{2}}_{\textsc{b}}\omega^{2}_{\textsc{b}}+e^{2}\ddot{\Gamma}^{(\textsc{e})}(0)\,.\label{E:irtyvgsd}
\end{align}
For a finite $\Lambda$, the values of $\ddot{\Gamma}^{(\textsc{e})}(0)$, $\dot{\Gamma}^{(\textsc{e})}(0)$, ${\Gamma}^{(\textsc{e})}(0)$ may not be humongous at all, so we may justifiably view the parameters in \eqref{E:irtyvgsd} as the effective ones, which have absorbed the corrections induced by the interaction between the system and the bath. Then the equation of motion \eqref{E:zrubgdh} becomes
\begin{align}\label{E:krhsgfsd}
	m_{\textsc{p}}\,\ddot{\chi}(t)+m^{\vphantom{2}}_{\textsc{p}}\omega^{2}_{\textsc{p}}-e^{2}\int_{0}^{t}\!ds\;\Gamma^{(\textsc{e})}(t-s)\,\dddot{\chi}(s)-e^{2}\Bigl[\ddot{\Gamma}^{(\textsc{e})}(t)\,\chi(0)+\dot{\Gamma}^{(\textsc{e})}(t)\,\dot{\chi}(0)+\Gamma^{(\textsc{e})}(t)\,\ddot{\chi}(0)\Bigr]=e\,\xi(t)\,,
\end{align}
in terms of physical or effective parameters. Note that $\dot{\Gamma}^{(\textsc{e})}(0)=0$ when the power of the monomial before $\mathcal{P}_{\Lambda}(\kappa)$ in the bath spectral density is an odd integer. {Here, the appearance of $\chi$ and its time derivatives at $t=0$ in the equation of motion {is a consequence of sudden switch-on and} casting the original equation of motion \eqref{E:zrubgdh} in terms of physical parameters like mass and natural frequency~\cite{CFdiss,Fleming}.} {On the other hand, if instead we introduce a gradual switching by explicitly implementing a time-dependent coupling constant $e(t)$ which increases sufficiently smoothly from the zero value at the initial time $t=0$ to full strength over a finite time interval, then the terms like $\chi$ and its time derivatives at $t=0$ in \eqref{E:krhsgfsd} may disappear, depending on the time derivatives of the coupling constant at $t=0$, as can be inspected by replacing $e\chi(s)$ with $e(s)\chi(s)$ in \eqref{E:dkhgfssg}.}

\subsection{Connection with the non-Markovian ALD equation}\label{S:jhvgdjdfg}

 {Before delving deeply into the issues associated with supra-Ohmic non-Markovian dynamics, let us take a quick look at how the non-Markovian version of ALD equation is related to \eqref{E:krhsgfsd} and what common features  they may share.}

 {We have stated in the previous section that from the viewpoint of non-Markovian dynamics, the dynamics described by the equation of motion \eqref{Ejhgdf} is destined to be unstable. In addition, we have come across quite a few subtleties and obscurities in interpreting this Markovian equation. The origin of these difficulties arises from the Markovian spectrum of the scalar field. This can be most clearly understood if we replace in  \eqref{Ejhgdf} the delta function kernel by a more general Ohmic non-Markovian kernel function $\Gamma^{(\textsc{e})}$, \eqref{E:wwww}. It also allows us to see the connection with the supra-Ohmic dynamics discussed earlier. Without repeating the derivation, we find that, from \eqref{Ejhgdf}, the non-Markovian ALD equation of motion corresponding to \eqref{E:irutyvs} takes the form
\begin{align}\label{E:rtgufyvdhf}
	m_{\textsc{p}}^{\vphantom{2}}\,\ddot{\chi}(t)+m_{\textsc{p}}^{\vphantom{2}}\omega_{\textsc{p}}^{2}\,\chi(t)&-e^{2}\int_{0}^{t}\!ds\;\Gamma^{(\textsc{e})}(t-s)\,\dddot{\chi}(s)-e^{2}\biggl[\Gamma^{(\textsc{e})}(t)\,\ddot{\chi}(0)+\dot{\Gamma}^{(\phi)}(t)\,\dot{\chi}(0)\biggr]=-e\,\dot{\phi}_{h}(\bm{x},t)\,,
\end{align}
with
\begin{align}\label{E:rkghbdf}
	m_{\textsc{p}}^{\vphantom{2}}&=m_{\textsc{b}}^{\vphantom{2}}+e^{2}\Gamma^{(\textsc{e})}(0)\,,&m_{\textsc{p}}^{\vphantom{2}}\omega_{\textsc{p}}^{2}&=m_{\textsc{b}}^{\vphantom{2}}\omega_{\textsc{b}}^{2}\,.
\end{align}
Comparing with \eqref{E:krhsgfsd}, we  readily see that \eqref{E:rtgufyvdhf} does not have the $\ddot{\Gamma}^{(\textsc{e})}(t)\,\chi(0)$ term, and the force terms on their righthand sides are not alike due to the different forms of the oscillator-field coupling.}

 {To see the relation of the non-Markovian ALD equation \eqref{Ejhgdf} with the supra-Ohmic equation of motion \eqref{E:zrubgdh}, we write the non-local term in \eqref{Ejhgdf} as
\begin{align}
	e^{2}\frac{\partial}{\partial t}\int_{0}^{t}\!ds\;G_{R,0}^{(\phi)}(t-s)\,\dot{\chi}(s)&=-e^{2}\ddot{\Gamma}^{(\textsc{e})}(0)\,\chi(t)+e^{2}\ddot{\Gamma}^{(\textsc{e})}(t)\,\chi(0)+e^{2}\int_{0}^{t}\!ds\;\partial^{3}_{s}\Gamma^{(\textsc{e})}(t-s)\,\chi(s)\,,
\end{align}
with the use of integration by parts and the substitution $G_{R,0}^{(\phi)}(\tau)=-\partial_{\tau}\Gamma^{(\textsc{e})}(\tau)$.  The third term on the right hand side gives \eqref{E:ruthit}, the whole nonlocal term in \eqref{E:zrubgdh}, but there are two additional terms in the non-Markovian ALD equation. The first term will account for additional frequency renormalization, not seen in \eqref{E:rkghbdf}, and the second term gives the contribution missing in \eqref{E:rtgufyvdhf} but present in \eqref{E:krhsgfsd}. Thus we see both systems have slightly different dependence on the initial conditions.}

 {The Laplace transform of the formal solution to \eqref{E:rtgufyvdhf} is given by
\begin{align}\label{E:nkbkcxjv}
	\tilde{\chi}(\sigma)&=\tilde{d}_{1}(\sigma)\,\chi(0)+\tilde{d}_{2}(\sigma)\,\dot{\chi}(0)-\frac{e}{m_{\textsc{p}}}\frac{\tilde{d}_{2}(\sigma)}{1-8\pi\gamma\,\sigma\Gamma^{(\textsc{e})}(0)}\,\bigl[\sigma\,\tilde{\phi}_{h}(\sigma)-\phi_{h}(0)\bigr]\,,
\end{align}
with
\begin{align}\label{E:kghdjtfs}
	\tilde{d}_{1}(\sigma)&=\frac{[1-8\pi\gamma\,\sigma\tilde{\Gamma}^{(\textsc{e})}(\sigma)]\sigma}{\sigma^{2}+\omega^{2}_{\textsc{p}}-8\pi\gamma\,\sigma^{3}\tilde{\Gamma}^{(\phi)}(\textsc{e})}\,,&\tilde{d}_{2}(\sigma)&=\frac{1-8\pi\gamma\,\sigma\Gamma^{(\textsc{e})}(0)}{\sigma^{2}+\omega^{2}_{\textsc{p}}-8\pi\gamma\,\sigma^{3}\tilde{\Gamma}^{(\textsc{e})}(\sigma)}\,.
\end{align}
We see that in a more general non-Markovian setting, even though the equation of motion has terms that depend on $\dddot{\chi}(t)$ and the initial condition $\ddot{\chi}(0)$, its solution actually depends only on two initial conditions $\chi(0)$ and $\dot{\chi}(0)$, thus supporting our earlier interpretation of the delta function in deriving \eqref{E:kgjdbf}. Taking a closer look at \eqref{E:rtgufyvdhf}, we notice that when we perform the Laplace transformation, the nonlocal term in \eqref{E:rtgufyvdhf} will yield a contribution to cancel the $\ddot{\chi}(0)$ term. Since \eqref{E:kghdjtfs} has the same denominator in $\tilde{d}_{i}(\sigma)$ as \eqref{E:ritbdghsd} below, both sets of $\tilde{d}_{i}(\sigma)$ will be subjected to the same criterion of stability, and have the same relaxation dynamics.}

 {In summary, compared with the generic supra-Ohmic non-Markovian dynamics described by \eqref{E:zrubgdh}, the non-Markovian ALD equation is shown to have different dependence on the initial conditions, but they have the same relaxation dynamics and stability criterion. These features of  the generic supra-Ohmic non-Markovian dynamics we shall discuss below will equally apply to the non-Markovian ALD equation. At the end of this section, we will revisit the Markovian limit of \eqref{E:rtgufyvdhf}.}

\subsection{Higher derivative terms in the equation of motion}
To better understand this supra-Ohmic Langevin equation we comment on some of its uncommon characteristics. Superficially, Eq.~\eqref{E:irtyvgsd} contains a third-order time derivative $\dddot{\chi}(t)$ of $\chi(t)$ and depends explicitly on the initial condition $\ddot{\chi}(0)$. These two features are not seen in the typical equations of motion, and are often considered as signaling something unphysical. They also create difficulties in implementing numerical solutions of this equation because it seems that an additional initial condition $\ddot{\chi}(0)$ is needed as well. However, this does not really pose any problem. When we carry out the Laplace transformation of \eqref{E:krhsgfsd} to find the Laplace transform $\tilde{\chi}(\sigma)$,  we obtain
\begin{align}\label{E:kgbhkfg}
	\tilde{\chi}(\sigma)=\tilde{d}_{1}(\sigma)\,\chi(0)+\tilde{d}_{2}(\sigma)\,\dot{\chi}(0)+\frac{e}{m_{\textsc{p}}}\frac{\tilde{d}_{2}(\sigma)}{1-8\pi\gamma\Gamma^{(\textsc{e})}(0)}\,\tilde{\xi}(\sigma)\,,
\end{align}
with
\begin{align}\label{E:ritbdghsd}
	\tilde{d}_{1}(\sigma)&=\frac{[1-8\pi\gamma\Gamma^{(\textsc{e})}(0)]\,\sigma}{\sigma^{2}+\omega_{\textsc{p}}^{2}-8\pi\gamma\tilde{\Gamma}^{(\textsc{e})}(\sigma)\,\sigma^{3}}\,,&\tilde{d}_{2}(\sigma)&=\frac{1-8\pi\gamma\Gamma^{(\textsc{e})}(0)}{\sigma^{2}+\omega_{\textsc{p}}^{2}-8\pi\gamma\tilde{\Gamma}^{(\textsc{e})}(\sigma)\,\sigma^{3}}\,.
\end{align}
The inverse Laplace transformations of $\tilde{d}_{1}(\sigma)$ and $\tilde{d}_{2}(\sigma)$ give a special set of fundamental solutions that {in the time domain} satisfy
\begin{align}
	d_{1}(0)&=1\,,&\dot{d}_{1}(0)&=0\,,&d_{2}(0)&=0\,,&\dot{d}_{2}(0)&=1\,,
\end{align}
by construction. In particular $\dfrac{1}{m_{\textsc{p}}}\dfrac{d_{2}(\tau)}{1-8\pi\gamma\Gamma^{(\textsc{e})}(0)}$ can be identified as the retarded Green's function $G_{R}^{(\chi)}(\tau)$ of $\chi$ associated with the equation of motion, \eqref{E:krhsgfsd}.

Eq.~\eqref{E:kgbhkfg} does not depend on $\ddot{\chi}(0)$, so the appearance of $\ddot{\chi}(0)$ in \eqref{E:krhsgfsd} is nothing but superficial. The Laplace transform of the nonlocal term will yield an additional $\ddot{\chi}(0)$ to cancel with the corresponding term in the second line of \eqref{E:krhsgfsd}. Thus, we only actually need two initial conditions $\chi(0)$ and $\dot{\chi}(0)$, instead of three. More importantly, we emphasize that even when we deal with the nonlocal equation of motion, we still need only \text{two} initial conditions, instead of the initial history, commonly met for the delayed differential equation. {The latter is often seen in the context of spatial non-Markovianity due to the case, for example, when the oscillator is in the proximity of a boundary~\cite{HL08} or when the constituents of the system is spatially separated~\cite{HH16}}.

\subsection{Runaway behavior of solutions}
The presence of $\dddot{\chi}(t)$ in the equation of motion \eqref{E:krhsgfsd} does not always imply a runaway result. Here in particular from \eqref{E:ritbdghsd}, we note that its solution depends on the factor $1-8\pi\gamma\Gamma^{(\textsc{e})}(0)$. The extra term $-8\pi\gamma\Gamma^{(\textsc{e})}(0)$ arrises from $-e^{2}\,\dot{\Gamma}^{(\textsc{e})}(t)\,\dot{\chi}(0)$ in the second line of \eqref{E:krhsgfsd}, and this type of term is not seen in the Ohmic case because we only apply integration by parts once on the corresponding nonlocal term {in the Ohmic case}. The factor $1-8\pi\gamma\Gamma^{(\textsc{e})}(0)$, though odd looking, is necessary to ensure the proper normalization of $d_{1}(t)$ and $d_{2}(t)$ such that the conditions $d_{1}(0)=1$ and $\dot{d}_{2}(0)=1$ are satisfied. In turn, these ensure the correct behavior of the nonequilibrium Robertson-Schr\"odinger function {at early times}.

Furthermore, this factor has a special significance with regard to the dynamical stability of this class of supra-Ohmic, non-Markovian Langevin equation \eqref{E:zrubgdh} under consideration. When this factor is positive, the fundamental solution $d_{1}(t)$ and $d_{2}(t)$ in general exhibit damping behavior, at least for the classes of bath spectral densities we are considering. On the other hand, if this factor is negative, then the solution increases without bound. To understand why the dynamics is unstable when $1-8\pi\gamma\,\Gamma^{(\textsc{e})}(0)<0$, let us examine the denominator of $\tilde{d}_{2}(\sigma)$, $\tilde{\mathfrak{J}}(\sigma)=\sigma^{2}+\omega_{\textsc{p}}^{2}-8\pi\gamma\,\tilde{\Gamma}^{(\textsc{e})}(\sigma)\,\sigma^{3}$, along the real $\sigma$ axis. Suppose $\tilde{\Gamma}^{(\textsc{e})}(\sigma)$ is well defined in the limit $\sigma\to0$. Thus we find $\displaystyle\lim_{\sigma\to0}\tilde{\mathfrak{J}}(\sigma)=\omega_{\textsc{p}}^{2}>0$. On the other hand $\displaystyle\lim_{\sigma\to\infty}\tilde{\Gamma}^{(\textsc{e})}(\sigma)$ in the $\sigma$ domain corresponds to $\displaystyle\lim_{t\to0}\Gamma^{(\textsc{e})}(0)\,\theta(t)$ in the $t$ domain. Thus we have
\begin{equation}\label{E:hgjdhgd}
	\lim_{\sigma\to\infty}\tilde{\Gamma}^{(\textsc{e})}(\sigma)\sim\pounds\Gamma^{(\textsc{e})}(0)\,\theta(t)=\frac{\Gamma^{(\textsc{e})}(0)}{\sigma}\,,
\end{equation}
so that in the limit $\sigma\to\infty$, we have 
\begin{equation}
		\lim_{\sigma\to\infty}\tilde{\mathfrak{J}}(\sigma)\simeq\bigl[1-8\pi\gamma\,\Gamma^{(\textsc{e})}(0)\bigr]\sigma^{2}+\mathcal{O}(\sigma)\,,
\end{equation}
going unbounded quadratically with $\sigma$. When $1<8\pi\gamma\,\Gamma^{(\textsc{e})}(0)$, we find $\tilde{\mathfrak{J}}(\sigma)<0$ in the $\sigma\to\infty$ limit. Now we obtain $\displaystyle\lim_{\sigma\to0}\tilde{\mathfrak{J}}(\sigma)>0$ but $\displaystyle\lim_{\sigma\to\infty}\tilde{\mathfrak{J}}(\sigma)<0$, we conclude that $\tilde{\mathfrak{J}}(\sigma)$ must have at least one real root between $\sigma=0$ and $\sigma\to\infty$. That is, $\tilde{d}_{2}(\sigma)$ will have a real positive pole. This will cause the runaway behavior in the solution to the equation of motion \eqref{E:krhsgfsd}.

Following the previous analysis, we observe that the presence of the factor $1-8\pi\gamma\,\Gamma^{(\textsc{e})}(0)$ will make $\tilde{d}_{2}(\sigma)$ assume the form $+1/\sigma^{2}$ as $\sigma\to\infty$. This implies that as $t>0$, $d_{2}(t)$ will grow like $+t$. The plus sign here is physically necessary. Suppose we have the initial condition $\chi(0)=0$ but $\dot{\chi}(0)\neq0$, then around the initial time, in the absence of any external force, the solution $\chi(t)$ is roughly given by $d_{2}(t)\,\dot{\chi}(0)$. Physically, we expect that in the absence of external forces, if the system is initially and momentarily at rest, then it should be displaced in the same direction as the initial velocity. It implies $d_{2}(t)$ must take on positive values around the initial time. It would be odd if it moves in a direction opposite to the initial speed. Thus the factor $1-8\pi\gamma\,\Gamma^{(\textsc{e})}(0)$ ensures physical consistency. Finally we mentioned that the nonequilibrium fluctuation-dissipation relation of the system, valid when $1>8\pi\gamma\,\Gamma^{(\textsc{e})}(0)$, still takes on the standard expression $\bar{G}_{H}^{(\chi)}(\kappa)=\coth\dfrac{\beta\kappa}{2}\,\operatorname{Im}\bar{G}_{R}^{(\chi)}(\kappa)$ after the system reaches its final equilibrium state. Its form is not affected by the presence of the factor.

Since the cutoff scale $\Lambda$ is contained in $\Gamma^{(\textsc{e})}(0)$, it implies that there exists a critical value $\Lambda_{c}$ such that when $\Lambda>\Lambda_{c}$, the supra-Ohmic non-Markovian system we consider here becomes unstable. When phrased in terms of the memory time, if the memory time $\Lambda^{-1}$ is shorter than the critical time scale $\Lambda_{c}^{-1}$, the system exhibits runaway behavior. Therefore, the Markovian limit of this supra-Ohmic system is purely formal, and in principle does not exist physically. Later when we take a closer look at the effects of various bath spectral density, we will give a more vivid demonstration of this important point.

The previous discussion reveals that the existence of the $\dddot{\chi}(t)$ in the equation of motion of a supra-Ohmic system does not automatically imply the necessity of additional initial condition, nor dynamical instability of the system. The latter primarily hinges on the duration of the memory time of the bath. {This modus operandi should work equally well in the Markovian limit. However this is not the case in the conventional treatment of the non-relativistic ALD equation (see discussion in Sec.~\ref{S:bgjvcher}). The aforementioned consistency consideration thus motivates our interpretation of the delta function arising from taking the Markovian limit in terms of its asymptotic form which supports our analysis expounded in Appendix~\ref{S:rjubgdfg}.}

Finally, we write down the expression for {the dispersion $\langle p^{2}(t)\rangle$ of the canonical momentum $p=m\dot{\chi}$ conjugated to $\chi$,} 
\begin{equation}
	\langle p^{2}(t)\rangle=m^{2}\dot{d}_{1}^{2}(t)\,\langle\chi^{2}(0)\rangle+\dot{d}_{2}^{2}(t)\,\langle p^{2}(0)\rangle+\frac{e^{2}}{[1-8\pi\gamma\,\sigma\Gamma(0)]^{2}}\int_{0}^{t}\!ds\,ds'\;\dot{d}_{2}(t-s)\,\dot{d}_{2}(t-s')\,G_{H,0}^{(\textsc{e})}(s,s')\,,
\end{equation}
because we need its late times $t\to\infty$ limit,
\begin{align}
	\langle p^{2}(\infty)\rangle=\frac{m_{\textsc{p}}}{1-8\pi\gamma\,\Gamma^{(\textsc{e})}(0)}\,\operatorname{Im}\int_{-\infty}^{\infty}\!\frac{d\kappa}{2\pi}\;\kappa^{2}\coth\frac{\beta\kappa}{2}\,\bar{d}_{2}(\kappa)=m_{\textsc{p}}\operatorname{Im}\int_{-\infty}^{\infty}\!\frac{d\kappa}{2\pi}\;\kappa^{2}\coth\frac{\beta\kappa}{2}\,\bar{G}_{R}^{(\chi)}(\kappa)\,,
\end{align}
with the Fourier transform $\bar{d}_{2}(\kappa)$ of $d_{2}(t)$ given by $\bar{d}_{2}(\kappa)=\tilde{d}_{2}(-i\kappa)$, to explore the effect of the bath spectrum. Here we have used the identity
\begin{align}
	\operatorname{Im}\bar{d}_{2}(\kappa)=\frac{\bar{d}_{2}^{\vphantom{*}}(\kappa)-\bar{d}_{2}^{*}(\kappa)}{2i}=8\pi\gamma\,\bigl[1-8\pi\gamma\,\Gamma^{(\textsc{e})}(0)\bigr]^{-1}\lvert\bar{d}_{2}(\kappa)\rvert^{2}\operatorname{Im}\bar{G}_{R,0}^{(\textsc{e})}(\kappa)\,,
\end{align}
to simplify the reuslt. Thus $\langle p^{2}(\infty)\rangle$ takes the same form as in the Ohmic case because the factor $1-8\pi\gamma\,\Gamma^{(\textsc{e})}(0)$ is canceled.

\subsection{Markovian limit of the non-Markovian ALD equation}
 {If we formally take the Markovian limit of the non-Markovian ALD equation  \eqref{E:rtgufyvdhf}, even though the limit does not meaningfully exist, by substituting $\Gamma^{(\textsc{e})}(t)$ with $\dfrac{1}{2\pi}\,\delta(t)$, we see that Eq.~\eqref{E:rtgufyvdhf} reduces to \eqref{E:irutyvs}. Furthermore, following our conventions, if we substitute
\begin{align}
	\tilde{\Gamma}^{(\textsc{e})}(\sigma)&=\frac{1}{2\pi}\tilde{\delta}(\sigma)\,,&&\text{with}&\tilde{\delta}(\sigma)&=\frac{1}{2}\,,&\tilde{\dot{\delta}}(\sigma)&=-\delta(0)+\frac{\sigma}{2}
\end{align}
into \eqref{E:nkbkcxjv}, we will consistently obtain \eqref{E:kgjdbf}. This shows that our algorithm of treating the integrals containing the delta function indeed gives a consistent formal Markovian limit from the full non-Markovian dynamics of the charged particle coupled to the electromagnetic fields.}

Therefore we infer that the dynamics of a point charge, when coupled to an electromagnetic field, is destined to be unstable from the viewpoint of non-Markovian dynamics, based on the connection with the supra-Ohmic system \eqref{E:zrubgdh} we established in Sec.~\ref{S:jhvgdjdfg}, because even though their equations of motion are slightly dissimilar and the solutions have different dependence on the initial conditions,  they have the same denominator in the respective Laplace transforms $\tilde{d}_{2}(\sigma)$, and thus the same pole structure. The locations of its poles determine the dynamical stability of both systems. {Stable dynamics is possible only when the cutoff scale in the frequency spectrum of the electromagnetic field is smaller than a critical value, which may be identified with the inverse of the classical charge radius. On the other hand, the  dependence on two (and only two) initial conditions relies on the consistent treatment of the Markovian limit, {in compliance with the relevant physical conditions.}

In this section we have discussed the generic features of a supra-Ohmic, non-Markovian system described by \eqref{E:zrubgdh},  {and its connection with the dynamics of the charged particle in the electromagnetic field. In the next section, we choose the bath's spectral density in the Lorentz-Drude form as examples for illustration. In Appendix~\ref{S:erbsdkf} we offer two more examples of bath spectral densities  commonly met - the exponentially decaying form and the hard cutoff form. We will highlight peculiar behavior of the fundamental solution associated with the latter bath spectral density.}

\section{Memory in the bath spectrum and the `runaway' issue}\label{S:tubgyidf}

In this section we  investigate the `runaway' issue in radiation reaction dynamics with an example. As remarked earlier, a finite cutoff frequency in the bath introduces a time scale which renders the system's dynamics non-Markovian. {The cutoff frequency of the bath determines the relaxation rate of the system and when the system's dynamics becomes unstable. Here we discuss the Lorentzian spectral function as a commonly used representative of supra-Ohmic baths. Discussions of two more examples will be given in Appendix~\ref{S:erbsdkf}. }

When the bath spectrum takes on the Lorentzian form
\begin{equation}\label{E:poqnzqq}
	J(\kappa)=\kappa^{3}\,\frac{\Lambda^{2}}{\kappa^{2}+\Lambda^{2}}\,,
\end{equation}
we find that the kernel function $\Gamma^{(\textsc{e})}(\tau)$ and its derivatives are given by
\begin{align}
	\Gamma^{(\textsc{e})}(\tau)&=\int_{-\infty}^{\infty}\!\frac{d\kappa}{2\pi}\;\frac{I(\kappa)}{2\pi\kappa}\,e^{-i\kappa\tau}=\frac{\Lambda}{4\pi}\,e^{-\Lambda\lvert\tau\rvert}\,,&&\Rightarrow&\Gamma^{(\textsc{e})}(0)&=\frac{\Lambda}{4\pi}\,,\\
	\dot{\Gamma}^{(\textsc{e})}(\tau)&=-\frac{\Lambda^{2}}{4\pi}\,e^{-\Lambda\lvert\tau\rvert}\,\operatorname{sgn}(\tau)\,,&&\Rightarrow&\dot{\Gamma}^{(\textsc{e})}(0)&=0\,,\\
	\ddot{\Gamma}^{(\textsc{e})}(\tau)&=\frac{\Lambda^{3}}{4\pi}\,e^{-\Lambda\lvert\tau\rvert}\,\operatorname{sgn}^{2}(\tau)-\frac{\Lambda^{2}}{2\pi}\,\delta(\tau)\,,&&\Rightarrow&\dot{\Gamma}^{(\textsc{e})}(0)&=-\frac{\Lambda^{2}}{2\pi}\,\delta(0)\,,
\end{align}
where we have used convention $\operatorname{sgn}^{2}(0)=0$, not 1, and
\begin{equation}
	\tilde{\Gamma}^{(\textsc{e})}(\sigma)=\frac{\Lambda}{4\pi}\frac{1}{\sigma+\Lambda}\,,
\end{equation}
such that the Laplace transforms of the fundamental solutions are
\begin{align}
	\tilde{d}_{1}(\sigma)&=(1-2\gamma\Lambda)\,\frac{\sigma}{\sigma^{2}+\omega_{\textsc{p}}^{2}-\frac{2\gamma\Lambda}{\sigma+\Lambda}\,\sigma^{3}}\,,\label{E:fbghjbd1}\\
	\tilde{d}_{2}(\sigma)&=(1-2\gamma\Lambda)\,\frac{1}{\sigma^{2}+\omega_{\textsc{p}}^{2}-\frac{2\gamma\Lambda}{\sigma+\Lambda}\,\sigma^{3}}\,.\label{E:fbghjbd2}
\end{align}
It tells us that the critical value of $\Lambda$ is given by $\Lambda_{c}=\frac{1}{2\gamma}$. Since $2\gamma=\dfrac{e^{2}}{4\pi m}$, the parameter $2\gamma$ plays the role similar to the ``classical charge radius'' in electrodynamics. That is,  when the memory time $\Lambda^{-1}$ is approximately this ``classical charge radius'' $2\gamma$, the behavior of the oscillator undergoes a transition from a stable regime to an unstable regime with runaway solutions. Before we proceed to explain its physical meaning, we first examine the $\Lambda$ dependence of the poles of $\tilde{d}_{2}(\sigma)$. As is shown in Fig.~\ref{Fi:supraPolEv}, the real parts of the poles of $\tilde{d}_{2}(\sigma)$ are not always negative, as is the case in the Ohmic bath, when we vary $\Lambda$. When $\Lambda<\Lambda_{c}$, the fundamental solution $d_{2}(t)$, the inverse Laplace transform of $\tilde{d}_{2}(\sigma)$, exhibits oscillatory but decaying behavior with time. On the other hand, when $\Lambda>\Lambda_{c}$, the presence of the positive real parts of the poles of $\tilde{d}_{2}(\sigma)$ implies exponential growth of $d_{2}(t)$.  This is related to the `runaway solutions' issue in the description of a classical point charge coupled to a electromagnetic field. The memory time can be understood as the time to cross the radius of effective influence, and is related to the finite effective size of the oscillator. Here we approach this issue from the perspective of memories in non-Markovian dynamics. Similar observations can  also be made from the perspective of  spatial nonlocality such as assuming a finite size for the charge or mass greater than a threshold value. Finite size effect in this context is discussed often in the literature \cite{GH3,Yaghjian,Erber,Pena,GLR}.

\begin{figure}
\centering
    \scalebox{0.5}{\includegraphics{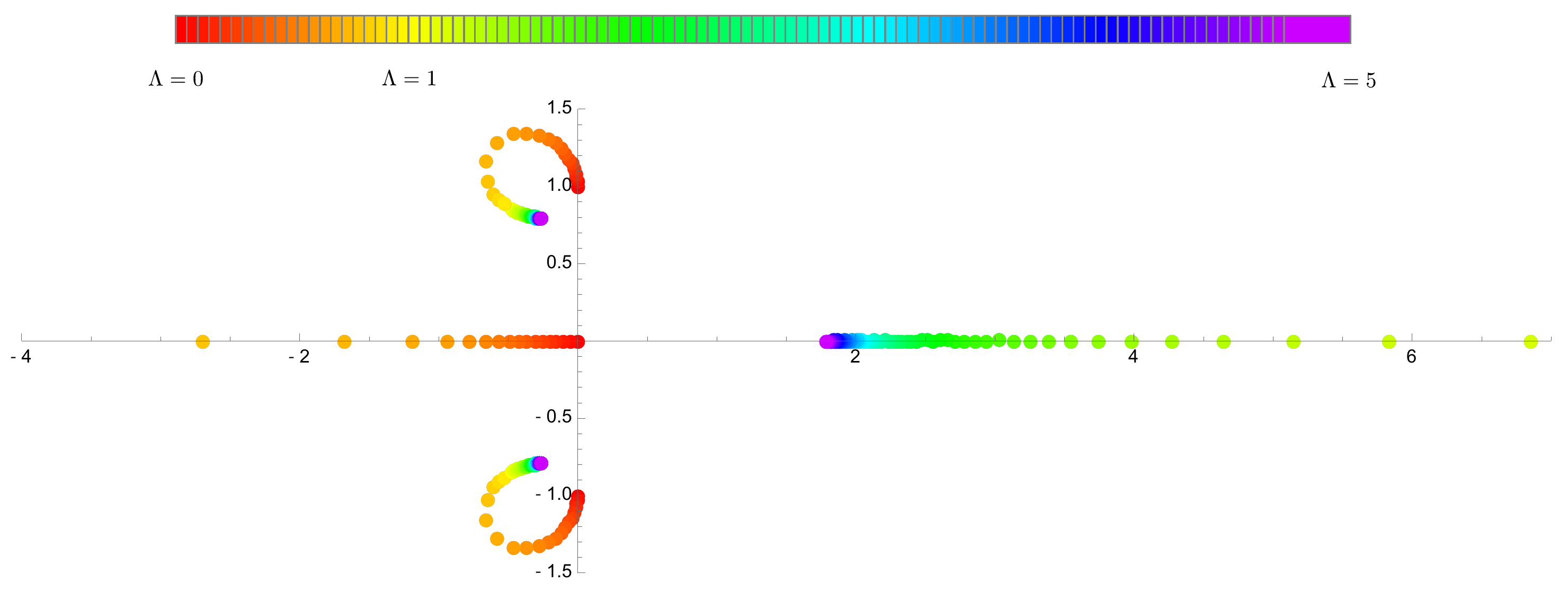}}
    \caption{The $\Lambda$ dependence of the poles of $\tilde{d}_{2}(\sigma)$, based on the equation of motion \eqref{E:zrubgdh} and the spectral model of the bath \eqref{E:poqnzqq}, on the complex $\sigma$ plane. The parameter $\Lambda$ varies from 0 to 5, in the unit of $\omega_{\textsc{p}}$, visualized by different hues from red, orange, yellow, green, blue, purple to magenta. We choose $\gamma=0.5\times\omega_{\textsc{p}}^{-1}$. The real parts of the poles can be positive when the value of $\Lambda$ is greater than the threshold value $\Lambda_{c}=(2\gamma)^{-1}=1$. It then implies the existence of the runaway behavior.}\label{Fi:supraPolEv}
\end{figure}

In Fig.~\ref{Fi:supraPolEv} we also note that the value of the positive real pole decreases to a limiting point\footnote{it is roughly 1.46557 for our choice of parameter. more precisely it given by
\begin{align}
	\sigma_{\star}&=\frac{1}{6\gamma}\bigl[1+\Xi^{\frac{1}{3}}+\Xi^{-\frac{1}{3}}\bigr]\,,&\Xi&=1+54\eta^{2}-6\eta\sqrt{3+81\eta^{2}}\,,&\eta&=\gamma\omega_{\textsc{p}}\,.
\end{align}
} with increasing $\Lambda$. On the other hand, for the regime of stable motion, the real part of the pole has the trend that it becomes more negative with the larger value of $\Lambda$, unless in the vicinity of $\Lambda_{c}$. Thus it tells us that if the memory time is longer, then the relaxation time scale is likewise longer. Both results seem to imply a strange phenomenon. When $\Lambda$ is slightly smaller than the critical value $\Lambda_{c}$, the system experiences very strong damping. In contrast, when $\Lambda$ is marginally great than $\Lambda_{c}$, the system has the most dramatic exponential-growth rate. The culprit of such drastic behavior lies in the fact at $\Lambda=\Lambda_{c}$, the denominator of $\tilde{d}_{2}(\sigma)$ is degenerate and reduces to the polynomial 
\begin{equation}
	\frac{\sigma^{2}}{1+2\gamma\sigma}+\omega_{\textsc{p}}^{2}\,,
\end{equation}
which has only two roots, instead of three
\begin{equation}
	-\gamma\omega_{\textsc{p}}^{2}\pm i\,\omega_{\textsc{p}}\sqrt{1-\gamma^{2}\omega_{\textsc{p}}^{2}}\,.
\end{equation}
These two roots have the negative real parts, so the system is still stable when $\Lambda=\Lambda_{c}$. Trivially $\Lambda=0$ is another case where the denominator of $\tilde{d}_{2}(\sigma)$ has only two pure imaginary roots. That is, effectively the system does not couple to the bath.

\begin{figure}
\centering
    \scalebox{0.35}{\includegraphics{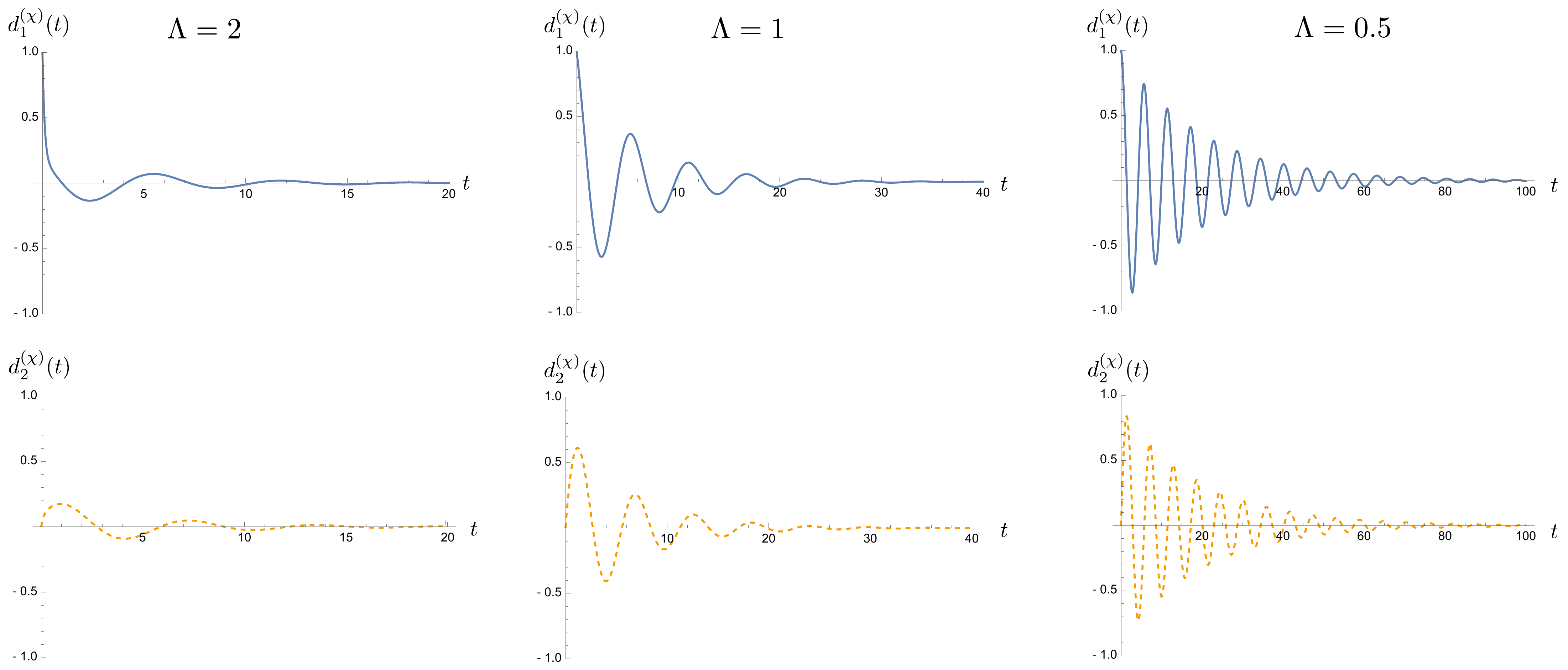}}
    \caption{The time evolution of the fundamental solutions for three different values 2, 1, 0.5 of $\Lambda$, when the bath spectral function takes the Lorentzian form, \eqref{E:poqnzqq}. The blue solid curve corresponds to $d_{1}^{(\chi)}(t)$, the orange dashed curve to {$d_{2}^{(\chi)}(t)$}. The time is expressed in the units of $\omega_{\textsc{p}}^{-1}$. These curve are found by taking the inverse Laplace transformations of \eqref{E:fbghjbd1} and \eqref{E:fbghjbd2}. Here we choose $\gamma=0.1$ and $\omega_{\textsc{p}}=1$, so $(\gamma\omega_{\textsc{p}}^{2})^{-1}=10$.}\label{Fi:Lorentzian}
\end{figure}

To make more explicitly the influence of the cutoff scale on the relaxation time, we show in Fig.~\ref{Fi:Lorentzian}, the temporal behavior of the fundamental solutions $d_{1}(t)$ and $d_{2}(t)$ for a few choices of $\Lambda$. The effective relaxation time  {$\gamma_{\textsc{eff}}^{-1}$, given below in \eqref{E:qdkjbgs}, is roughly related to $\Lambda^{-2}$, the memory time  {squared}. It is quite different from the value $(\gamma\omega_{\textsc{p}}^{2})^{-1}$ we would expect in the Markovian limit, for example, as in the dynamics of the charge oscillator, treated by order reduction.} Thus the coherent superposition due to longer memory inhibits the damping. The memory time $\Lambda^{-1}$ determined how long the past history of the oscillator can affect its present dynamics. The ratio $\omega_{\textsc{p}}/\Lambda$ dictates the importance of  the memory effect. When the ratio is smaller than unity, the memory time is short compared to the typical time scale of motion, so the memory effect does not add up to be significant. However, when the ratio is greater than or equal to unity, the motion in the previous cycle can be coherently added up to the present cycle. This feature is particularly prominent when the ratio is much greater than one, so that the motion at the current moment is superposed by the corresponding copies in many previous cycles.

Quantitatively, when memory is present and sustains longer than a critical period $\Lambda_{c}^{-1}$, the supra-Ohmic system will relax with a time scale $\gamma_{\textsc{eff}}^{-1}$. For sufficiently small $\Lambda$, as shown in Fig.~\ref{Fi:suprOhmNonMark}, {the effective damping constant} $\gamma_{\textsc{eff}}$ is given by
\begin{equation}\label{E:qdkjbgs}
	\gamma_{\textsc{eff}}\sim\gamma\Lambda^2\biggl(\frac{1}{1-2\gamma\Lambda}-\frac{\Lambda^2}{\omega_{\textsc{p}}^2+\Lambda^2}\biggr)\,.
\end{equation}
This is much smaller than $\gamma\omega_{\textsc{p}}^{2}$. Thus, the non-Markovian effect very effectively slows down the relaxation of the oscillator's dynamics, such that it essentially behaves like a system weakly coupled to the bath. However, this does not necessarily imply the the weak-coupling Markovian description is a good substitute of the strong-coupling non-Markovian system because the former may not sufficiently accurately get hold of the phase information inherent in the latter~\cite{HAH22}.

\begin{figure}
\centering
    \scalebox{0.35}{\includegraphics{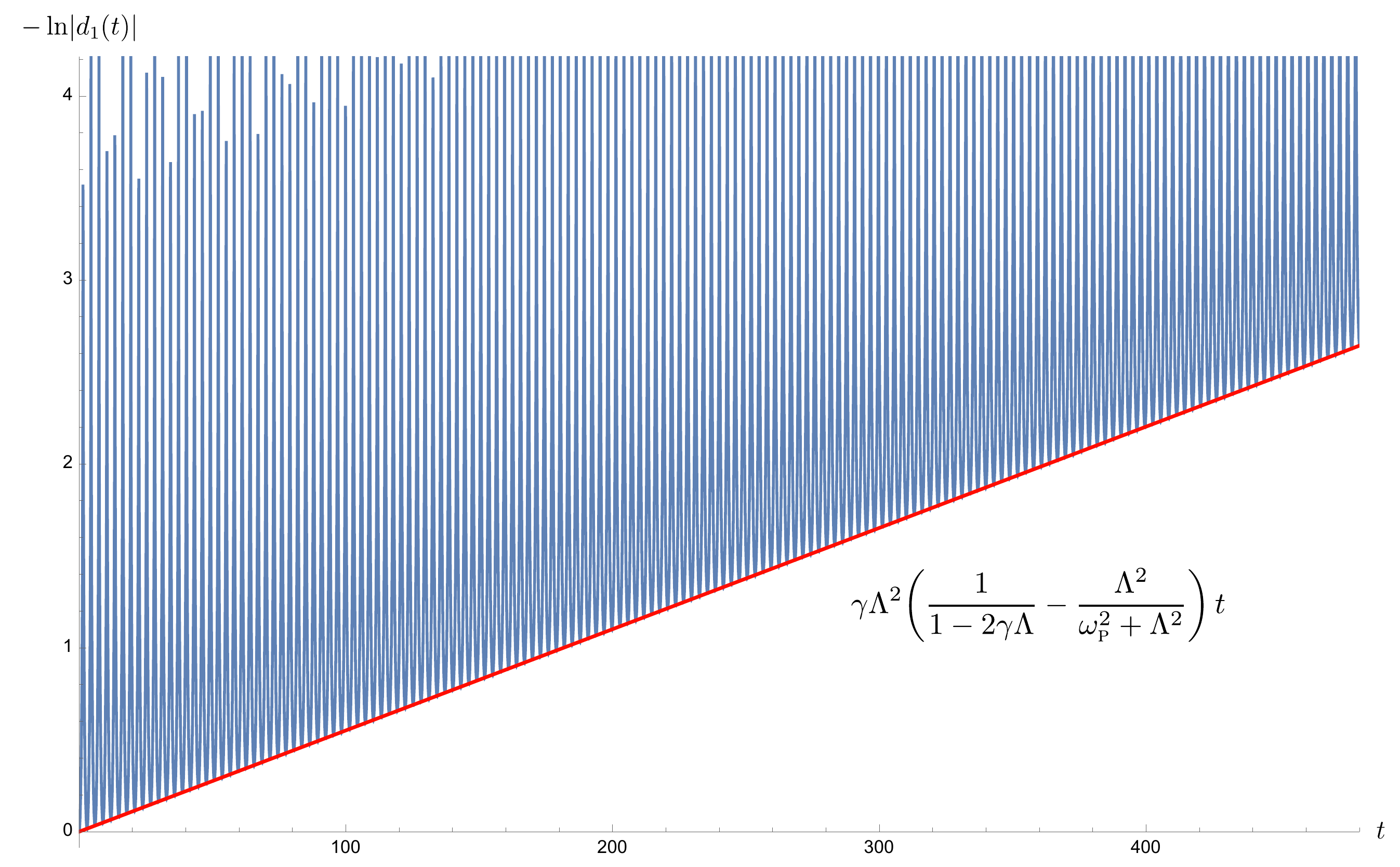}}
    \caption{The temporal evolution of $-\ln d_{1}(t)$ for $\Lambda=0.1\times\omega_{\textsc{p}}$ and $\gamma=0.5\times\omega_{\textsc{p}}$, described by the equation of motion \eqref{E:zrubgdh} and the spectral model of the bath \eqref{E:poqnzqq}. The horizontal $t$ axis takes the unit of $\omega_{\textsc{p}}^{-1}$. The lower bound of $-\ln d_{1}(t)$ is approximately tangent to $\gamma_{\textsc{eff}}t$, where $\gamma_{\textsc{eff}}$ is given in \eqref{E:qdkjbgs}.}\label{Fi:suprOhmNonMark}
\end{figure}

\section{Summary and Conclusion}

The ALD equation, which describes the dynamics of a point charge in the electromagnetic field, contains a third-order time derivative of the charge's position. Its presence implies a few  unwelcome features such as the need for additional initial conditions,  runaway solutions, pre-acceleration, with causality violation consequences. In fact, in the open-system framework where the charge is the system of interest and the electromagnetic field serve as its environment, the ALD equation can be understood as the equation of motion in the Markovian limit of the reduced system  from a fully non-Markovian dynamical equation which contains a few more terms in the form of products of delta function or its time derivative and the initial conditions, when they are ignored at $t>0$ after the interaction is switched on at $t=0$.

We advocate taking this broader open systems perspective in treating backreaction problems where memories associated with multiple  time-scales from the environment impact directly on the non-Markovian dynamics of the reduced system.  From this perspective we conclude that the ALD equation should not be viewed  as a standalone equation but the Markovian limit of a class of supra-Ohmic, non-Markovian dynamics obeying a third-order integro-differential equation. The salient features are listed below: 
\begin{enumerate}
	\item The supra-Ohmic non-Markovian dynamics in general is not {always} stable and {may} not have a well-defined Markovian limit even if the corresponding Ohmic, non-Markovian counterpart has these nice properties.
	\item The dynamical stability of the supra-Ohmic, non-Markovian equation depends on the cutoff scale, or the length of the memory time. When the memory time is shorter than a critical value, the system  {may} transit from the stable regime to the runaway regime.
	\item Although the equation of motion contains a third-order time derivative, its solution still depend on two initial conditions, instead of three.
	\item The presence of the third-order time derivative in the supra-Ohmic, non-Markovian equation of motion thus is not to blame for the runaway dynamics and the need of additional initial condition. 
\end{enumerate}

From these aspects, we found that in deriving the equation of motion of a charged particle coupled to an electromagnetic field, those initial-condition dependent terms that have been dropped to obtain the ALD equation have to be retained because 1) they are the essential ingredients to address the initial condition problem and 2) they are needed to approach a consistent Markovian limit. More importantly, when we take the (formal) Markovian limit by pushing the cutoff scale to infinity, we will encounter highly singular expressions. They appear when the nonlocal kernel of the bath reduces to a delta function in such a limit.  Thus,  3) to maintain a consistent Markovian limit, conforming to physical configurations, we should  interpret the emergent delta function and its derivatives in terms of their asymptotic forms. In so doing, the ambiguity of treating the delta functions at the initial time can be resolved. These practices ensure that the solution to the equation of motion of the charged particle depends on only two initial conditions}.

The runaway behavior, on the other hand, is inevitable in this Markovian limit because the system is memoryless. The supra-Ohmic system is doomed to be unstable when the memory time is shorter than a critical value. We learn that the instability of the ALD equation is a consequence of taking the Markovian limit, not the presence of third-order time derivatives in the equation.

{In conclusion, in a fully non-Markovian formulation, using the non-Markovian ALD equation \eqref{E:rtgufyvdhf}, one can achieve a pathology-free description of the supra-Ohmic dynamics, with the distinct advantages that a) it requires only two, not three, specified initial conditions,  and b) it ensures stable dynamics without runaway and pre-acceleration when the memory time is greater than the critical value.}\\

\noindent

{\bf Acknowledgment} We thank Onat Ar{\i}soy for discussions on the late time behavior of non-Markovian dynamics for some of the bath spectra. J.-T. Hsiang is supported by the Ministry of Science and Technology of Taiwan, R.O.C. under Grant No.~MOST 111-2811-M-008-022.

\appendix
\section{Ohmic dynamics}\label{S:oetruihisurt}
Here we review the dynamics of a Brownian oscillator, coupled to a free Ohmic bath. This will serve as a comparison with the supra-Ohmic case.

The spectral density $I(\kappa)$ of the bath will assume the form $\kappa\,\mathcal{P}_{\Lambda}(\kappa)$, where $\mathcal{P}_{\Lambda}(\kappa)$ falls off to zero for $\kappa$ greater than the cutoff scale $\Lambda$, which is the only scale in $\mathcal{P}_{\Lambda}(\kappa)$, to suppress the contributions from the high frequency modes of the bath, and $\mathcal{P}_{\Lambda}(\kappa)=1$ as $\Lambda\to\infty$.

The retarded Green's function $G_{R,0}^{(\textsc{e})}(\tau)$ and the Hadamard functions $G_{H,0}^{(\textsc{e})}(\tau)$ associated with this Ohmic bath are given by
\begin{align}
	G_{R,0}^{(\textsc{e})}(\tau)&=\theta(\tau)\,G_{P,0}^{(\textsc{e})}(\tau)=i\,\theta(\tau)\int_{0}^{\infty}\!\frac{d\kappa}{2\pi}\;\frac{I(\kappa)}{2\pi}\,\Bigl[e^{-i\kappa\tau}-e^{+i\kappa\tau}\Bigr]\,,\\
	G_{H,0}^{(\textsc{e})}(\tau)&=\int_{-\infty}^{\infty}\!\frac{d\kappa}{2\pi}\;\frac{I(\kappa)}{4\pi}\coth\frac{\beta\kappa}{2}\,e^{-i\kappa\tau}\,,
\end{align}
where $G_{p,0}^{(\textsc{e})}(\tau)$ is the Pauli-Jordan function and $\beta$ is the inverse initial bath temperature. It proves convenient to  introduce a new kernel function $\Gamma^{(\textsc{e})}(\tau)$ by
\begin{equation}\label{E:gfjbser}
	G_{P,0}^{(\textsc{e})}(\tau)=-\frac{\partial}{\partial\tau}\Gamma^{(\textsc{e})}(\tau)\,,
\end{equation}
so that its Fourier representation is given by
\begin{equation}\label{E:ebdgghfs}
	\Gamma^{(\textsc{e})}(\tau)=\int_{-\infty}^{\infty}\!\frac{d\kappa}{2\pi}\;\frac{I(\kappa)}{2\pi\kappa}\,e^{-i\kappa\tau}\,.
\end{equation}

The equation of motion for the displacement of the Brownian oscillator takes the form
\begin{equation}\label{E:kdfgjbhdf}
	m_{\textsc{b}}^{\vphantom{2}}\,\ddot{\chi}(t)+m_{\textsc{b}}^{\vphantom{2}}\omega_{\textsc{b}}^{2}\chi(t)-e^{2}\int_{0}^{t}\!ds\;G_{R,0}^{(\textsc{e})}(t-s)\chi(s)=e\,\phi_{h}(t)\,,
\end{equation}
when the oscillator's displacement $\chi(t)$ is bilinearly coupled to the bath variable $\phi(t)$ with the coupling strength $e$. The driving force $\phi_{h}(t)$ denotes the free bath component, excluding the backreaction from the oscillator, which is accounted for by the nonlocal expression in \eqref{E:kdfgjbhdf}. The parameters $m_{\textsc{b}}$, $\omega_{\textsc{b}}$ are undetermined for the moment, and will be renormalized according to the oscillator-bath interaction. This interaction is switched on instantaneously at $t=0$, and the states of these two sub-systems are assumed to be initially uncorrelated. The overhead dot on a quantity represents taking the derivative of it with respect to time $t$.

Putting \eqref{E:gfjbser} into \eqref{E:kdfgjbhdf} allows us to identify the contribution to the parameter's renormalization,
\begin{align}\label{E:rofbdgf}
	m_{\textsc{b}}^{\vphantom{2}}\,\ddot{\chi}(t)+\Bigl[m_{\textsc{b}}^{\vphantom{2}}\omega_{\textsc{b}}^{2}-e^{2}\Gamma^{(\textsc{e})}(0)\Bigr]\,\chi(t)+e^{2}\Gamma^{(\textsc{e})}(t)\,\chi(0)+e^{2}\int_{0}^{t}\!ds\;\Gamma^{(\textsc{e})}(t-s)\dot{\chi}(s)=e\,\phi_{h}(t)\,.
\end{align}
Only the natural frequency $\omega_{\textsc{b}}$ needs renormalizing, so we assign
\begin{align}
	m_{\textsc{p}}&=m_{\textsc{b}}\,,&{\omega_{\textsc{p}}^{2}}&={\omega_{\textsc{b}}^{2}}-\frac{e^{2}}{m_{\textsc{p}}}\,\Gamma^{(\textsc{e})}(0)\,.
\end{align}
At first it looks odd that the initial condition $\chi(0)$ appears in the equation of motion. This is the consequence when we try to explicitly isolate the component that contributes to the parameter renormalization/correction. The formal solution can be conveniently found with the help of the Laplace transformation,
\begin{equation}\label{E:dkfvjfd}
	\tilde{\chi}(\sigma)=\tilde{d}_{1}(\sigma)\,\chi(0)+\tilde{d}_{2}(\sigma)\,\dot{\chi}(0)+\cdots\,,
\end{equation}
where we have conveniently introduced $e^{2}=8\pi\gamma m_{\textsc{p}}$, and $\cdots$ represents the particular solution that depends on $\phi_{h}$. The meaning of the damping constant $\gamma$ will be made clear later. Two transforms $\tilde{d}_{1}(\sigma)$, $\tilde{d}_{2}(\sigma)$ in \eqref{E:dkfvjfd} take the form
\begin{align}
	\tilde{d}_{1}(\sigma)&=\frac{\sigma}{\sigma^{2}+\omega_{\textsc{p}}^{2}+8\pi\gamma\,\sigma\,\tilde{\Gamma}^{(\textsc{e})}(\sigma)}\,,&\tilde{d}_{2}(\sigma)&=\frac{1}{\sigma^{2}+\omega_{\textsc{p}}^{2}+8\pi\gamma\,\sigma\,\tilde{\Gamma}^{(\textsc{e})}(\sigma)}\,.
\end{align}
In the current case their inverse transforms $d_{1}(t)$ and $d_{2}(t)$ are related by $d_{1}(t)=\dot{d}_{2}(t)$, but it may not always be the case, so we will keep the notation $d_{1}(t)$. Here we note that unlike the supra-Ohmic case in \eqref{E:ritbdghsd}, they do not have the prefactor that is related to $\Gamma^{(\textsc{e})}(0)$. Thus following the analysis that leads to \eqref{E:hgjdhgd}, the Ohmic dynamics tends to have a well-defined Markovian limit, at least for the bath spectral densities we have considered. In other words, even if the considered Ohmic, non-Markovian dynamics has a well-defined Markovian limit, this does not guarantee that the corresponding supra-Ohmic, non-Markovian dynamics has a meaningful Markovian limit.

Suppose the cutoff frequency $\Lambda$ is the only scale in $\mathcal{P}_{\Lambda}(\kappa)$. Then its inverse $\Lambda^{-1}$ gives the duration in which $\Gamma^{(\textsc{e})}(t)$ does not drop off to zero. In other word, according to the integral expression in \eqref{E:rofbdgf}, the parameter $\Lambda^{-1}$ roughly tells how long the dynamics of $\chi(t)$ depends on its previous history. Thus $\Lambda^{-1}$ can be interpreted as a memory time with the Markovian limit $\Lambda\to\infty$ understood as being memoryless. The dynamics of $\chi(t)$ is purely local, independent of the state of motion in earlier moments. When $\tilde{d}_{2}(\sigma)$ does not have any pole for any value of $\Lambda$ at the $\operatorname{Re}\sigma>0$ half of the complex $\sigma$ plane, the dynamics is always stable and the Markovian limit exists. For example, since $\tilde{\Gamma}^{(\textsc{e})}(\sigma)$ can be found to be
\begin{equation}
	\tilde{\Gamma}^{(\textsc{e})}(\sigma)=\int_{0}^{\infty}\!d\tau\;e^{-\sigma\tau}\!\int_{-\infty}^{\infty}\!\frac{d\kappa}{2\pi}\;\frac{I(\kappa)}{2\pi\kappa}\,e^{-i\kappa\tau}=\frac{1}{2\pi i}\int_{-\infty}^{\infty}\!\frac{d\kappa}{2\pi}\;\frac{I(\kappa)}{\kappa(\kappa-i\sigma)}\,,
\end{equation}
if $\mathcal{P}_{\Lambda}(\kappa)$ takes a Lorentzian shape
\begin{equation}
	\mathcal{P}_{\Lambda}(\kappa)=\frac{\Lambda^{2}}{\Lambda^{2}+\kappa^{2}}\,,
\end{equation}
then we find 
\begin{equation}\label{E:eotsbdg}
	\tilde{\Gamma}^{(\textsc{e})}(\sigma)=\frac{1}{4\pi}\frac{\Lambda}{\Lambda+\sigma}\,,
\end{equation}
and its Markovian limit $\Lambda\to\infty$ gives $\dfrac{1}{4\pi}$. In the latter case, the transforms $\tilde{d}_{1}(\sigma)$, $\tilde{d}_{2}(\sigma)$ have the standard forms for the damped oscillator, with $\gamma$ serving as the damping constant.

\begin{figure}
    \centering
    \scalebox{0.35}{\includegraphics{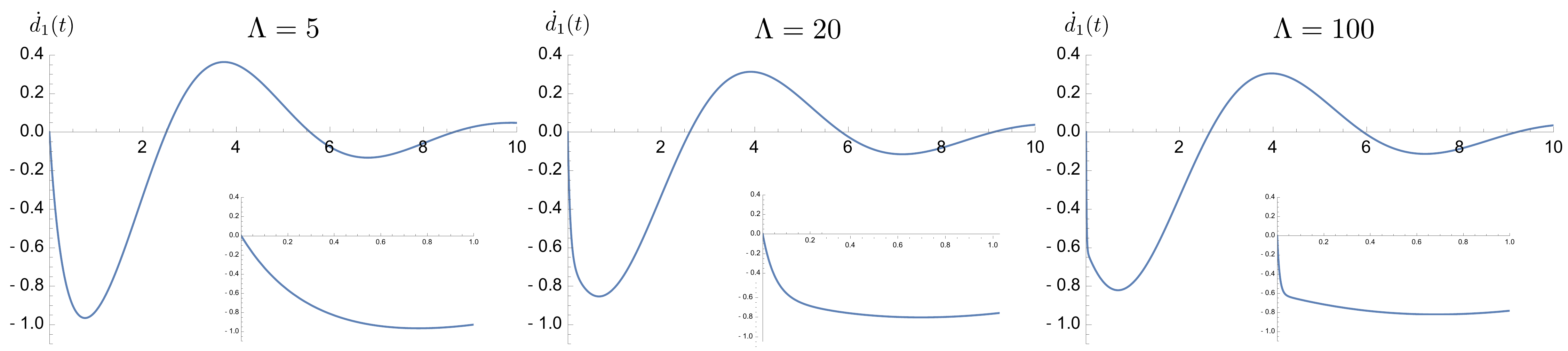}}
    \caption{Here we show the behavior of $\dot{d}_1(t)$ with time for different choices of the cutoff scale $\Lambda$, and particularly in the inset, we blow up the region $0\leq t\leq1$ to reveal the effect of the cutoff. We express the parameters in the unit of $\omega_\textsc{p}$, and choose $\gamma=0.3$, $\omega_\textsc{p}=1$.}\label{Fi:dotd1}
\end{figure}
Here in fact we will see a minor ambiguity arising from the Markovian limit, where $\tilde{d}_{1}(\sigma)$ is given by
\begin{equation}    
    \tilde{d}_{1}(\sigma)=\frac{\sigma}{\sigma^{2}+\omega_{\textsc{p}}^{2}+2\gamma\,\sigma}\,.
\end{equation}
Carrying out the inverse Laplace transformation gives
\begin{equation}\label{E:tyudgfg}
    d_1(t)=e^{-\gamma t}\cos\Omega t-\frac{\gamma}{\Omega}\,e^{-\gamma t}\sin\Omega t\,,
\end{equation}
where $\Omega=\sqrt{\omega^2_{\textsc{p}}-\gamma^2}$. We immediately note that $d_1(0)=1$, but $\dot{d}_1(0)=-2\gamma\neq0$, contradicting the requirements about the fundamental solution, $d_1(0)=1$ and $\dot{d}_1(0)=0$. This can be resolved once we inspect the equation of motion in this limit
\begin{equation}
    \ddot{\chi}(t)+2\gamma\,\dot{\chi}(t)+\omega_{\textsc{p}}^2\chi(t)+4\gamma\,\delta(t)\chi(0)=0\,,
\end{equation}
where we have ignored the force $e\,\phi_h$(t) because it is irrelevant to the discussion. If we move $4\gamma\,\delta(t)\chi(0)$ to the righthand side, and treat it as a driving force, then we see that this term offers a kick at $t=0$ and instantaneously push $\dot{d}_1(t)$ from zero to $-2\gamma$ at $t=0^+$, causing the ``slip"~\cite{FRH11}. This treatment, though correct, is quite awkward. If we start from \eqref{E:rofbdgf} and, for example, \eqref{E:eotsbdg} for a finite cutoff $\Lambda$, we will see that $\dot{d}_1(t)$ indeed starts from 0 smoothly. Nonetheless, when we increase $\Lambda$, the behavior of $\dot{d}_1(t)$ about $t=0$ shows a progressively rapid but smooth drop toward $-2\gamma$ within the time scale $\Lambda^{-1}$, as shown in Fig.~\ref{Fi:dotd1}. The inexplicable behavior of \eqref{E:tyudgfg} is then nicely clarified.

This is the most elementary example in Brownian motion showing the difficulties arising from treating the equations as Markovian ab initio.

\section{Ambiguities associated with the delta function in the integral transformation}\label{S:rjubgdfg}
As is well-known,  the Dirac delta function $\delta(t)$ is a distribution, not a function, possessing these `peculiar' properties
\begin{align}\label{E:rfbgdke}
	1)\qquad\delta(t)&=0\,,\qquad\text{when $t\neq0$}\,,&&2)\qquad\int_{-\infty}^{\infty}\!dt\;\delta(t)=1\,,\\
	&&&3)\qquad\int_{-\infty}^{\infty}\!dt\;\dot{\delta}(t)\,f(t)=-{\dot{f}(0)}\,.
\end{align}
Thus we often resort to its asymptotic forms to explore its properties in various contexts. Two common asymptotic forms are
\begin{align}\label{E:troidfn}
	\delta_{\epsilon}(t)&=\frac{1}{2\epsilon}\Bigl[\theta(t+\epsilon)-\theta(t-\epsilon)\Bigr]\,,&\delta_{\epsilon}(t)&=\frac{1}{\sqrt{2\pi\epsilon}}\,e^{-\frac{t^{2}}{2\epsilon}}\,,
\end{align}
with $\epsilon$  {being positive and infinitesimally small}. The former has finite support but is not continuous, while the latter is a smooth Gaussian with width $\epsilon$. When $\epsilon$ approaches zero, both become very sharp and asymptotically satisfy the properties in \eqref{E:rfbgdke}. Here we attempt to set up a consistent and unified modus operandi xsuitable for our purpose of taking the Markovian limit of the non-Markovian dynamics of the Brownian oscillator. Namely, we only view the delta function as a convenient substitute used in the sensible Markovian limit. Otherwise, the asymptotic form of the delta function is probably more compatible with the physical settings\footnote{We thank Daniel Reiche for discussions on this point}.

When the delta function is involved in an integral representation, ambiguities may arise when the ``peak'' of the delta function is located at the integration limits. This happens frequently in the context of taking the Markovian limit. There, we often come across an expression like
\begin{equation}
	I_{0}(t)=\int_{0}^{t}\!ds\;\delta(t-s)\,f(s)\,,\label{E:peirhugsdf}
\end{equation}
for some well behaved ordinary function $f(s)$ and $t\geq0$. Our convention assigns the value $\dfrac{1}{2}f(t)$, instead of $f(t)$, because heuristically only half of the delta function ``peak'' is used in the integral. However, strictly speaking, {when we set} $t=0$, the integral gives indefinite answers, 0 or $f(0)/2$, depending on how lax  we are with the interpretation of the delta function. We will return to this point below.

Another integral expression we often encounter is
\begin{equation}\label{E:ritugdf}
	I_{1}(t)=\int_{0}^{t}\!ds\;\partial_{s}\delta(t-s)\,f(s)\,.
\end{equation}
If we treat for the moment the delta function as an ordinary function, we will obtain
\begin{align}
	I_{1}(t)=\delta(0)\,f(t)-\delta(t)\,f(0)-\int_{0}^{t}\!ds\;\delta(t-s)\,\dot{f}(s)=\delta(0)\,f(t)-\delta(t)\,f(0)-\frac{1}{2}\,\dot{f}(t)\,,\label{E:uighsiufd}
\end{align}
after integration by parts. Comparing with \eqref{E:rfbgdke}, we cannot help wondering whether the first two terms on the righthand side of \eqref{E:uighsiufd} should be present. We use the first asymptotic form of the delta function in \eqref{E:troidfn} to examine \eqref{E:ritugdf}. Suppose for simplicity, we assume $f(t)=1+2t$, and obtain
\begin{align}
	\int_{0}^{t}\!ds\;\delta_{\epsilon}(t-s)\,\dot{f}(s)&=\theta(t)\theta(t-\epsilon)+\frac{t}{\epsilon}\Bigl[1-\theta(t)\theta(t-\epsilon)\Bigr]\,,\\
	\int_{0}^{t}\!ds\;\partial_{s}\delta_{\epsilon}(t-s)\,f(s)&=\frac{1+2t-2\epsilon}{2\epsilon}\,\theta(t)\theta(t-\epsilon)\,,
\end{align}
so that we have, for $\epsilon>0$,
\begin{align}
	\int_{0}^{t}\!ds\;\partial_{s}\delta_{\epsilon}(t-s)\,f(s)+\int_{0}^{t}\!ds\;\delta_{\epsilon}(t-s)\,\dot{f}(s)=\frac{1+2t}{2\epsilon}-\frac{1}{2\epsilon}\,\Bigl[1-\theta(t)\theta(t-\epsilon)\Bigr]\,.\label{E:kdgbkde}
\end{align}
Here we identify $\delta_{\epsilon}(0)=\dfrac{1}{2\epsilon}$. Mathematically this can be problematic because the $\epsilon\to0$ limit is required to be taken  {prior} to letting $t$ go to zero. Nonetheless we can be slightly lax about this. We have in mind that when we talk about the Markovian limit, we only require $\Lambda$ be sufficiently large, not necessarily infinite. Thus we will interpret the delta function by its asymptotic form.

In \eqref{E:kdgbkde}, we can write $1-\theta(t)\theta(t-\epsilon)$ into
\begin{align}
	1-\theta(t)\theta(t-\epsilon)=\Bigl[\theta(t)+\theta(-t)\Bigr]\,\theta(t+\epsilon)-\theta(t)\theta(t-\epsilon)=\theta(t)\Bigl[\theta(t+\epsilon)-\theta(t-\epsilon)\Bigr]\,,
\end{align}
when $t>0$. Thus, the righthand side of \eqref{E:kdgbkde} actually gives $\delta_{\epsilon}(0)\,f(t)-\delta_{\epsilon}(t)\,f(0)$, and hence we obtain \eqref{E:uighsiufd} in the asymptotic sense.

Finally, we look at the example
\begin{align}\label{E:guiddrt}
	I_{2}(t)=\partial_{t}\int_{0}^{t}\!ds\;\delta(t-s)\,f(s)\,.
\end{align}
By $I_{0}$ we may conclude that $I_{1}(t)$ is given by $\dfrac{1}{2}\dot{f}(t)$. However, when we carry out the differentiation first and treat for the moment the delta function as an ordinary function, we obtain
\begin{equation}
	I_{2}(t)=\delta(0)\,f(t)-\int_{0}^{t}\!ds\;\partial_{s}\delta(t-s)\,f(s)
\end{equation}
Using the result found in \eqref{E:uighsiufd}, we arrive at
\begin{equation}
	I_{2}(t)=\delta(t)\,f(0)+\dfrac{1}{2}\,\dot{f}(t)\,,
\end{equation}
and end up with
\begin{equation}
	\dfrac{1}{2}\,\dot{f}(t)\stackrel{?}{=}\delta(t)\,f(0)+\frac{1}{2}\,\dot{f}(t)\,,
\end{equation}
a paradoxical contradiction with \eqref{E:guiddrt} for $t\geq0$. It turns out that we should be more careful with $I_{0}(t)$, where we have implicitly assumed $t>0$. To take its time derivative, we need to consider the case $t<0$ as well. When $t<0$, $I_{0}(t)$ is given by
\begin{equation}
	I_{0}(t<0)=-\frac{1}{2}\,f(t)\,,
\end{equation}
and thus together with the $t\geq0$ result, we have
\begin{equation}\label{E:kfdbgdkf}
	I_{0}(t)=\frac{1}{2}\,f(t)\,\Bigl[\theta(t)-\theta(-t)\Bigr]\,,
\end{equation}
which has a jump at $t=0$. We also observe that though $\theta(0)$ is ambiguous, \eqref{E:kfdbgdkf} still gives a definite result 0 at $t=0$, thus extending the result in \eqref{E:peirhugsdf} to the whole $t$ axis. Then it implies that for $I_{2}(t)$, we have
\begin{equation}\label{E:itugiusrtse}
	I_{2}(t)=\delta(t)\,f(0)+\frac{1}{2}\,\dot{f}(t)\,\Bigl[\theta(t)-\theta(-t)\Bigr]\,,
\end{equation}
consistently.

Now we are ready to address the Laplace transformation of the delta function
\begin{equation}
	\tilde{\delta}(\sigma)=\int_{0}^{\infty}\!dt\;e^{-t\sigma}\,\delta(t)
\end{equation}
according to the definition of the Laplace transformation in \eqref{E:gbuess}. In the standard reference, the lower limit of the integral on the righthand side is chosen to be $0^{-}$ for $t>0$. Then by this convention we will have $\tilde{\delta}(\sigma)=1$. However, following the discussion around $I_{0}$, it seems more consistent to choose $\tilde{\delta}(\sigma)=1/2$. One one hand, this can be verified by the asymptotic form of the delta function, and on the other hand, if we would like to take the Markovian limit by smoothly changing the cutoff $\Lambda$ to infinity, it is the asymptotic form of the delta function that is relevant to our discussions, not the delta function itself per se. In this context, we view the latter as a mathematical convenience for the physical configuration.

Following the same spirit, by our convention, the Laplace transformation of $\dot{\delta}(t)$ is given by
\begin{equation}
	\int_{0}^{\infty}\!dt\;\dot{\delta}(t)\,e^{-t\sigma}=-\delta(0)+\frac{\sigma}{2}\,,
\end{equation}
instead of $\sigma$ given by the standard reference.

{Here we have discussed the algorithms of treating the delta function which arises from taking the Markovian limit of the non-Markovian dynamics. We interpret the delta function by its asymptotic form. In doing so, we are able to formulate the correct, unified and consistent way of taking the Markovian limit of non-Markovian dynamics.}

\section{more examples of the spectral densities}\label{S:erbsdkf}
Here we offer two more examples to illustrate the characteristics of supra-Ohmic non-Markovian dynamics in Sec.~\ref{S:fkgfdgfg}. The exponentially decaying bath spectral function in essence is complementary to the Lorentzian form \eqref{E:poqnzqq}, except that it has better suppression over the high-frequency modes of the bath. The hard-cutoff spectrum offers the most intuitive implementation of the cutoff scale, but it tends to introduce peculiar behavior in reduced non-Markovian dynamics, when the cutoff scale is not extremely large.

\subsection{Exponentially decaying spectrum}
When the bath spectrum has an exponentially decaying form
\begin{align}\label{E:rtbgdfhdf}
	J(\kappa)=\kappa^{3}\,e^{-\frac{\lvert\kappa\rvert}{\Lambda}}=\kappa^{2}\,I(\kappa)\,,
\end{align}
the $\Gamma^{(\textsc{e})}(\tau)$ kernel then has a Lorentzian form
\begin{align}
	\Gamma^{(\textsc{e})}(\tau)&=\int_{-\infty}^{\infty}\!\frac{d\kappa}{2\pi}\;\frac{I(\kappa)}{2\pi\kappa}\,e^{-i\kappa\tau}=\frac{\Lambda}{2\pi^{2}}\frac{1}{1+\Lambda^{2}\tau^{2}}\,,&&\Rightarrow&\Gamma^{(\textsc{e})}(0)&=\frac{\Lambda}{2\pi^{2}}\,,\\
	\dot{\Gamma}^{(\textsc{e})}(\tau)&=-\frac{\Lambda^{3}}{\pi^{2}}\frac{\tau}{(1+\Lambda^{2}\tau^{2})^{2}}\,,&&\Rightarrow&\dot{\Gamma}^{(\textsc{e})}(0)&=0\,,\\
	\ddot{\Gamma}^{(\textsc{e})}(\tau)&=-\frac{\Lambda^{3}}{\pi^{2}}\frac{1-3\Lambda^{2}\tau^{2}}{(1+\Lambda^{2}\tau^{2})^{3}}\,,&&\Rightarrow&\ddot{\Gamma}^{(\textsc{e})}(0)&=-\frac{\Lambda^{3}}{\pi^{2}}\,,
\end{align}
and its Laplace transform is given by
\begin{equation}
	\tilde{\Gamma}^{(\textsc{e})}(\sigma)=\frac{1}{4\pi}\,\cos\frac{\sigma}{\Lambda}+\frac{1}{2\pi^{2}}\biggl[\sin\frac{\sigma}{\Lambda}\,\operatorname{Ci}(\frac{\sigma}{\Lambda})-\cos\frac{\sigma}{\Lambda}\,\operatorname{Si}(\frac{\sigma}{\Lambda})\biggr]\,.
\end{equation}
From \eqref{E:ritbdghsd}, we then find the Laplace transforms of the fundamental solutions $d_{1}(t)$ and $d_{2}(t)$ are
\begin{align}
	\tilde{d}_{1}(\sigma)&=\Bigl(1-\frac{4\gamma\Lambda}{\pi}\Bigr)\frac{\sigma}{\sigma^{2}+\omega_{\textsc{p}}^{2}-\sigma^{3}\,\dfrac{4\gamma}{\pi}\,\biggl\{\dfrac{\pi}{2}\,\cos\dfrac{\sigma}{\Lambda}+\biggl[\sin\dfrac{\sigma}{\Lambda}\,\operatorname{Ci}(\dfrac{\sigma}{\Lambda})-\cos\dfrac{\sigma}{\Lambda}\,\operatorname{Si}(\dfrac{\sigma}{\Lambda})\biggr]\biggr\}}\,,\\
	\tilde{d}_{2}(\sigma)&=\Bigl(1-\frac{4\gamma\Lambda}{\pi}\Bigr)\frac{1}{\sigma^{2}+\omega_{\textsc{p}}^{2}-\sigma^{3}\,\dfrac{4\gamma}{\pi}\,\biggl\{\dfrac{\pi}{2}\,\cos\dfrac{\sigma}{\Lambda}+\biggl[\sin\dfrac{\sigma}{\Lambda}\,\operatorname{Ci}(\dfrac{\sigma}{\Lambda})-\cos\dfrac{\sigma}{\Lambda}\,\operatorname{Si}(\dfrac{\sigma}{\Lambda})\biggr]\biggr\}}
\end{align}
In this case the critical value is determined by $\Lambda_{c}=\frac{\pi}{4\gamma}$. However, since in this case the denominator of $\tilde{d}_{2}(\sigma)$ is a transcendental function of $\sigma$ and has a branchcut along $\operatorname{Re}\sigma<0$, it is much harder to identify the roots, and carry out the analysis like the previous Lorentzian case. However, the temporal behavior of the fundamental solutions share features similar to those in the Lorentzian case.

The exponentially decaying bath spectrum case has an advantage that the Lorentzian spectrum does not have. Let us choose a $\Lambda<\Lambda_{c}$ such that the system can be relaxed to an equilibrium state at sufficiently late times, greater than the effective relaxation time scale. Then we expect that the covariance matrix elements will approach to time-independent constants. Let us in particular look at the late time value of $\langle p^{2}(t)\rangle$, which is given by
\begin{align}
	\langle p^{2}(\infty)\rangle=\frac{m_{\textsc{p}}}{1-8\pi\gamma\,\Gamma^{(\textsc{e})}(0)}\,\operatorname{Im}\int_{-\infty}^{\infty}\!\frac{d\kappa}{2\pi}\;\kappa^{2}\coth\frac{\beta\kappa}{2}\,\bar{d}_{2}(\kappa)\,.\label{E:kfjgfd}
\end{align}
We observe that even the dynamics is stable when $\Lambda<\Lambda_{c}$, the late-time value of $\langle p^{2}(\infty)\rangle$ is not guaranteed to be finite. It depends on the power of $\kappa$ when $\lvert\kappa\rvert\to\infty$. The large $\kappa$ limit of $\operatorname{Im}\bar{d}_{2}(\kappa)$ has to be $\kappa^{\alpha}$ with $\alpha<-3$. When the bath spectrum takes the Lorentzian form, we find
\begin{equation}
	\lim_{\lvert\kappa\rvert\to\infty}\operatorname{Im}\bar{d}_{2}(\kappa)=\frac{2(1-\gamma\Lambda)\gamma\Lambda^{2}}{(1-2\gamma\Lambda)^{2}}\frac{1}{\kappa^{3}}+\mathcal{O}(\kappa^{-4})\,.
\end{equation}
In contrast, for the exponentially decaying bath spectrum, we instead have
\begin{equation}
	\lim_{\lvert\kappa\rvert\to\infty}\operatorname{Im}\bar{d}_{2}(\kappa)=e^{-\frac{\kappa}{\Lambda}}\Bigl(1-\frac{4\gamma\Lambda}{\pi}\Bigr)^{-1}\frac{4\gamma}{\kappa}+\mathcal{O}(\kappa^{-3})\,.
\end{equation}
Hence, we see that the Lorentzian bath spectrum does not suppress high-frequency bath modes enough to ease off the logarithmic divergence in the late-time value of $\langle p^{2}(t)\rangle$.

In passing, it is interesting to mention in this context the double Lorentzian bath spectrum
\begin{equation}\label{E:dkfhjdszqq}
	J(\kappa)=\kappa^{3}\,\Bigl(\frac{\Lambda^{2}}{\kappa^{2}+\Lambda^{2}}\Bigr)^{2}\,.
\end{equation}
It is then straightforward to show 
\begin{align}
	\tilde{d}_{1}(\sigma)&=\bigl(1-\gamma\Lambda\bigr)\frac{\sigma}{\sigma^{2}+\omega_{\textsc{p}}^{2}-\sigma^{3}\,\gamma\Lambda\,\dfrac{\sigma+2\Lambda}{(\sigma+\Lambda)^{2}}}\,,\\
	\tilde{d}_{2}(\sigma)&=\bigl(1-\gamma\Lambda\bigr)\frac{1}{\sigma^{2}+\omega_{\textsc{p}}^{2}-\sigma^{3}\,\gamma\Lambda\,\dfrac{\sigma+2\Lambda}{(\sigma+\Lambda)^{2}}}\,.
\end{align}
such that
\begin{equation}
	\lim_{\lvert\kappa\rvert\to\infty}\operatorname{Im}\bar{d}_{2}(\kappa)=\frac{2\gamma\Lambda^{4}}{(1-\gamma\Lambda)}\frac{1}{\kappa^{5}}+\mathcal{O}(\kappa^{-6})\,.
\end{equation}
Thus, although this bath spectrum algebraically falls off at large frequency, it provides sufficient suppression to kill off the logarithmic divergence in $\langle p^{2}(\infty)\rangle$. The {late-time values of the} other covariance matrix elements have lower powers of $\kappa$ in the corresponding integrands in \eqref{E:kfjgfd}, so they will have the sensible results without any divergence even for the Lorentzian bath spectrum.

\begin{figure}
\centering
    \scalebox{0.35}{\includegraphics{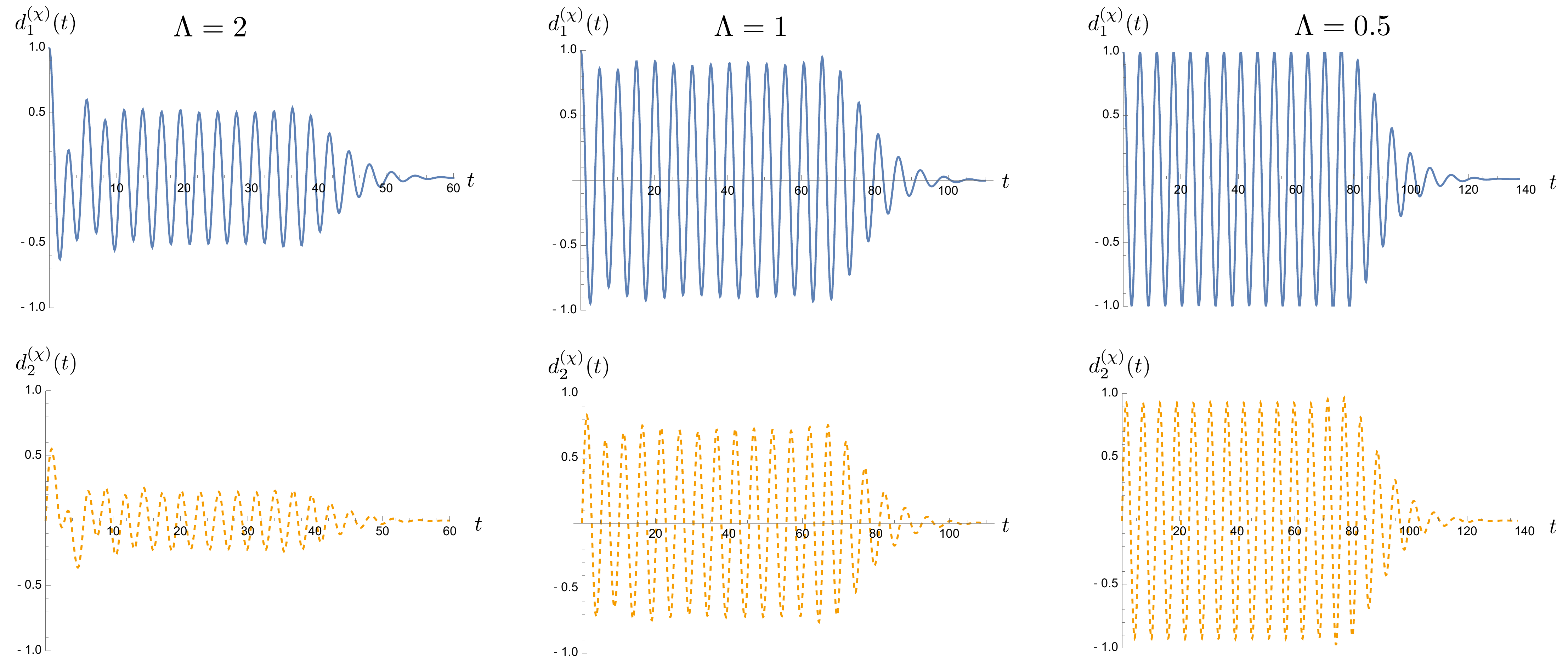}}
    \caption{The time evolution of the fundamental solutions when the bath spectral function takes the {hard-cutoff form \eqref{E:gdfbsa}}. The blue solid curve corresponds to $d_{1}^{(\chi)}(t)$, the orange dashed curve to {$d_{2}^{(\chi)}(t)$}. The time is expressed in the units of $\omega_{\textsc{p}}^{-1}$. From the left to the right, we choose three different values of $\Lambda$, which are 2, 1, 0.5. However, the decaying behavior is rather unusual compared with the other bath spectral densities we have chosen. The amplitude of oscillations is maintained over a time scale much greater than the typical relaxation time scale, even the memory time is short. For sufficiently small $\Lambda\lesssim1$, the effective relaxation time seems only mildly depends on it, so that the temporal behavior of the fundamental solutions for $\Lambda=1$ and $0.5$ does not change much, in strong contrast to the other two cases.}\label{Fi:hardCut}
\end{figure}

\subsection{Hard-cutoff spectrum}
Finally, we discuss the case when the bath spectrum is the band-limited form. Here we are interested in the specific form 
\begin{equation}\label{E:gdfbsa}
	J(\kappa)=\kappa^{3}\,\theta(\Lambda-\kappa)\theta(\kappa)\,,
\end{equation}
where $\Lambda$ is the cutoff frequency, beyond which the higher frequency modes are completely disregarded. Since we usually re-write the frequency integrals such that the integration range extends from $-\infty$ to $+\infty$, the spectral density \eqref{E:gdfbsa} is conveniently put into an alternative form
\begin{equation}\label{E:yeuiysbdf}
	J(\kappa)=\kappa^{3}\,\theta(\Lambda-\kappa)\theta(\kappa+\Lambda)\,.
\end{equation}

The corresponding kernel function $\Gamma^{(\textsc{e})}(\tau)$ takes the form
\begin{align}
	\Gamma^{(\textsc{e})}(\tau)&=\int_{-\infty}^{\infty}\!\frac{d\kappa}{2\pi}\;\frac{I(\kappa)}{2\pi\kappa}\,e^{-i\kappa\tau}=\int_{-\Lambda}^{\Lambda}\!\frac{d\kappa}{2\pi}\;\frac{1}{2\pi}\,e^{-i\kappa\tau}=\frac{1}{2\pi^{2}}\frac{\sin\Lambda\tau}{\tau}\,,\\
	\dot{\Gamma}^{(\textsc{e})}(\tau)&=\frac{\Lambda\tau\,\cos\Lambda\tau-\sin\Lambda\tau}{2\pi^{2}\tau^{2}}\,,\\
	\ddot{\Gamma}^{(\textsc{e})}(\tau)&=-\frac{\Lambda^{2}\tau^{2}\,\sin\Lambda\tau+2\Lambda\tau\,\cos\Lambda\tau-\sin\Lambda\tau}{2\pi^{2}\tau^{3}}\,,
\end{align}
and its Laplace transform
\begin{equation}
	\tilde{\Gamma}^{(\textsc{e})}(\sigma)=\frac{1}{2\pi^{2}}\,\tan^{-1}\frac{\Lambda}{\sigma}\,,
\end{equation}
so that
\begin{align}
	\tilde{d}_{1}(\sigma)&=\Bigl(1-\frac{4\gamma\Lambda}{\pi}\Bigr)\frac{\sigma}{\sigma^{2}+\omega_{\textsc{p}}^{2}-\sigma^{3}\dfrac{4\gamma}{\pi}\,\tan^{-1}\dfrac{\Lambda}{\sigma}}\,,\\
	\tilde{d}_{2}(\sigma)&=\Bigl(1-\frac{4\gamma\Lambda}{\pi}\Bigr)\frac{1}{\sigma^{2}+\omega_{\textsc{p}}^{2}-\sigma^{3}\dfrac{4\gamma}{\pi}\,\tan^{-1}\dfrac{\Lambda}{\sigma}}\,.
\end{align}
The transition occurs at $4\gamma\Lambda=\pi$, where the relative dominance of $\sigma^{2}$ and $\sigma^{3}\dfrac{4\gamma}{\pi}\,\tan^{-1}\dfrac{\Lambda}{\sigma}$ swaps for large $\sigma$. That is, the ultra-violet behavior of $\tilde{d}_{2}(\sigma)$ will depend on whether $4\gamma\Lambda\gtrless\pi$.

\begin{figure}
\centering
    \scalebox{0.33}{\includegraphics{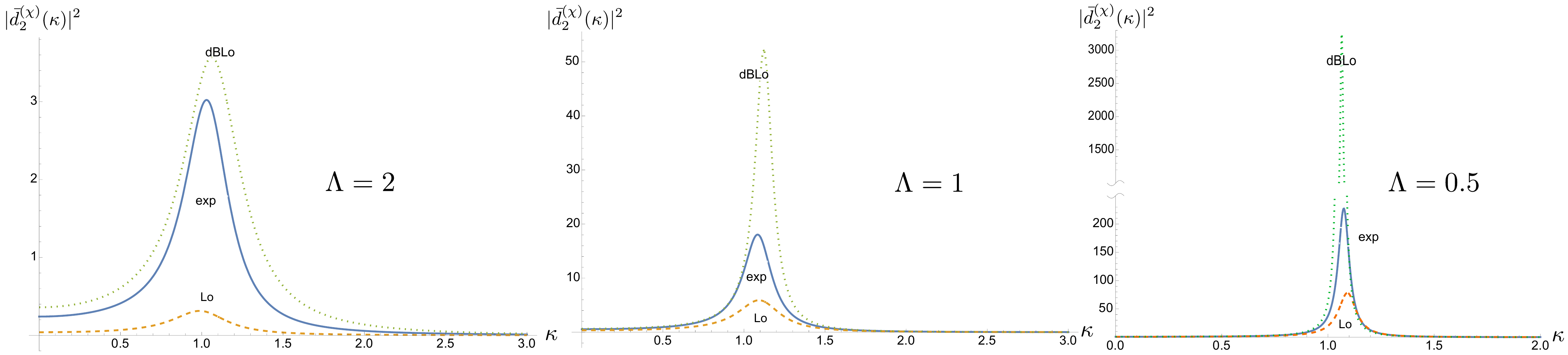}}
    \caption{The frequency response of the oscillator to various supra-Ohmic, non-Markovian bath spectra. Here we include that exponential decaying (exp) spectrum~\eqref{E:rtbgdfhdf}, the Lorentzian (Lo) spectrum~\eqref{E:poqnzqq} and the double Lorentzian (dBLo) spectrum~\eqref{E:dkfhjdszqq}. The height of the response curve reflects the strength of effective damping, i.e., effective relaxation time scale.}\label{Fi:freqRep}
\end{figure}

Compared with the previous two cases, the temporal evolution of the fundamental solution in the hard cutoff spectrum has an unusual feature. As seen in Fig.~\eqref{Fi:hardCut}, both $d_{1}(t)$ and $d_{2}(t)$ essentially oscillate with a constant amplitude for an extended period of time, and then rapidly decay to zero. They do not decay to zero roughly exponentially like the previous two cases. The period of constant-amplitude oscillations is correlated to the memory time $\Lambda^{-1}$.

To decrypt the underlying mechanism that yields such a distinctive nature, we will examine the response function $\lvert\bar{d}_{2}(\kappa)\rvert^{2}$ with $\bar{d}_{2}(\kappa)=\tilde{d}_{2}(-i\kappa)$, as we often do in the case of a driven, damped harmonic oscillator. In the latter case, the height of the resonance peak goes like $\gamma^{-2}$, so when   $\gamma\to0$ for a sinusoidally driven free oscillator, the response curve will become unbounded at the resonance frequency. Fig.~\ref{Fi:freqRep} shows the typical behavior of the response functions for the previous cases we have discussed. We see that the double Lorentzian spectrum induces a much weaker effective damping since the resonance peak is much sharper than the other spectral densities. For bath spectra considered there, we see the tendency that when the cutoff frequency $\Lambda$ gets smaller, the resonance peak is narrower and higher. These demonstrate that the effective damping strength decreases with a longer memory time, and the system behaves more like a weak-coupling system.

In contrast, the frequency response function in the hard cutoff case can have a positive, real pole. Thus it has an unbounded resonance peak, as in the case of the driven, undamped oscillator. The frequency response takes the form
\begin{equation}
	\lvert\bar{d}_{2}(\kappa)\rvert^{2}=\Bigl(1-\frac{4\gamma\Lambda}{\pi}\Bigr)^{2}\Bigl\{4\pi^{2}\gamma^{2}\kappa^{6}\,\Theta(\Lambda-\kappa)+\Bigl[\pi\kappa^{2}-\pi\omega_{\textsc{p}}^{2}+\gamma\kappa^{3}\ln\Bigl(\frac{\kappa-\Lambda}{\kappa+\Lambda}\Bigr)^{2}\Bigr]^{2}\Bigr\}^{-1}\,.\label{E:riugbfdg}
\end{equation}
When $\kappa<\Lambda$, the {denominator} in \eqref{E:riugbfdg} is always positive, so this pole is located at $\kappa>\Lambda$. It can be shown that the pole is given by the root of
\begin{equation}\label{E:odkkzsjak}
	\pi\kappa^{2}-\pi\omega_{\textsc{p}}^{2}+\gamma\kappa^{3}\ln\Bigl(\frac{\kappa-\Lambda}{\kappa+\Lambda}\Bigr)^{2}=0\,.
\end{equation}
This root has an interesting trend that when $\Lambda$ approaches $\Lambda_{c}=\frac{\pi}{4\gamma}$, the root will move to positive infinity. This can be shown from the $\kappa\to\infty$ limit of \eqref{E:odkkzsjak}
\begin{equation}
	\bigl(\pi-4\gamma\Lambda\bigr)\,\kappa^{2}+\mathcal{O}(\kappa^{0})=0\,.
\end{equation}
It is identically satisfied only if $\Lambda=\Lambda_{c}$.

\begin{figure}
\centering
    \scalebox{0.5}{\includegraphics{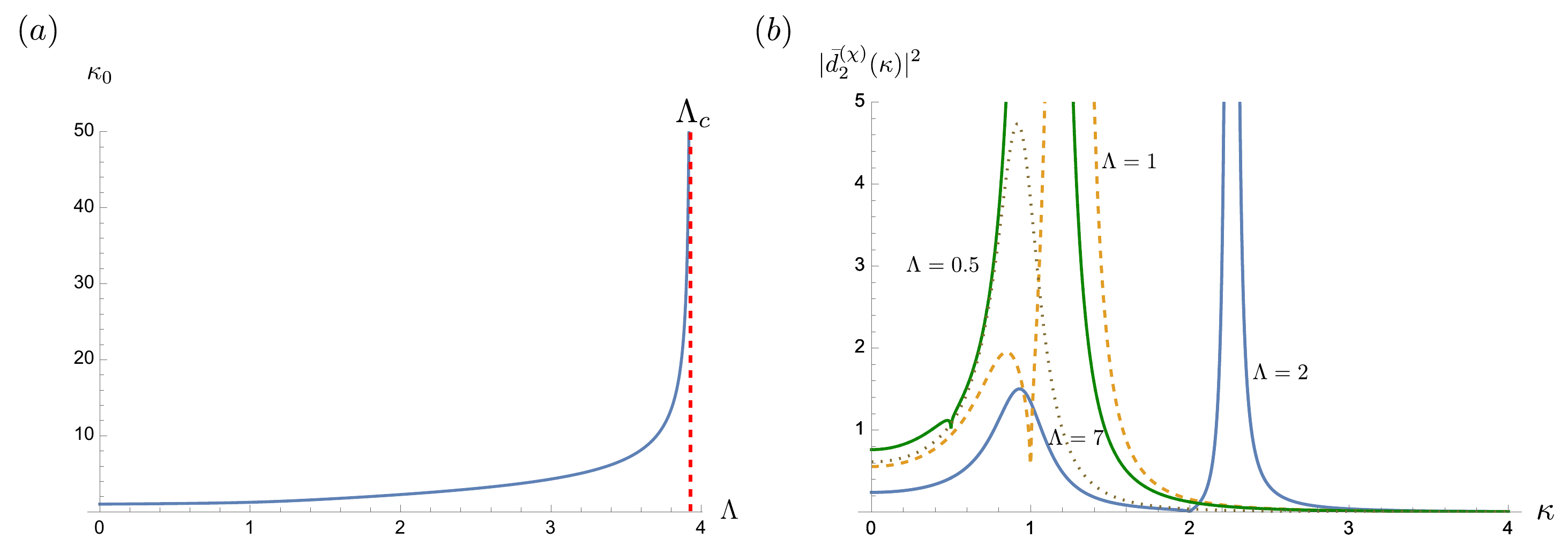}}
    \caption{(a) The dependence of the real pole $\kappa_{0}$ in \eqref{E:riugbfdg} on the cutoff $\Lambda$. The location of the pole moves to the infinity when $\Lambda$ approaches $\Lambda_{c}$. (b) The frequency response of the oscillator due to the hard cutoff bath spectrum. The green solid curve corresponds to $\Lambda=0.5$, the orange dashed curve $\Lambda=1$, and the blue solid curve is the $\Lambda=2$ case. When $\Lambda<\Lambda_{c}$, there are two resonance peaks, and one of them is unbounded. This unbounded peak disappears when $\Lambda>\Lambda_{c}$, for example, given by the dotted curve for $\Lambda=7$. Here we choose $\omega_{\textsc{p}}=1$, $\gamma=0.2$ so that $\Lambda_{c}=3.92699$ in the unit of $\omega_{\textsc{p}}$.}\label{Fi:HCpoles}
\end{figure}
There is bounded damped resonance peak at $\kappa<\Lambda$, and it is located at the root
\begin{equation}
	\frac{\partial}{\partial\kappa}\biggl\{4\pi^{2}\gamma^{2}\kappa^{6}+\Bigl[\pi\kappa^{2}-\pi\omega_{\textsc{p}}^{2}+\gamma\kappa^{3}\ln\Bigl(\frac{\kappa-\Lambda}{\kappa+\Lambda}\Bigr)^{2}\Bigr]^{2}\biggr\}=0\,.
\end{equation}
This is the only resonance peak we will have when $\Lambda>\Lambda_{c}$, that is, the Markovian limit. When $\Lambda>\omega_{\textsc{p}}$, the resonance peak is roughly located at $\kappa\sim\omega_{\textsc{p}}\bigl(1-3\gamma_{\vphantom{p}}^{2}\omega_{\textsc{c}}^{2}+\cdots\bigr)$, slightly smaller than $\omega_{\textsc{p}}$. Thus the location and the shape of this bounded peak is expected to be nudged by the unbounded peak when $\Lambda$ moves across $\omega_{\textsc{p}}$. Fig.~\ref{Fi:HCpoles}--(b) shows that the feature of the bounded peak becomes vanishingly small in the small $\Lambda$ limit, say $\Lambda=0.5$, very different from the other cases. It implies that when the memory time is long, the damping becomes ineffective, but by a different mechanism. This probably explains why the fundamental solutions of the oscillator, when coupled to the bath that has a hard cutoff spectrum, can have a nearly constant oscillation amplitude over a prolonged length of time and then decay. When $\Lambda<\Lambda_{c}$, the former is related to the unbounded resonance, while the latter is caused by the bounded resonance peak. When $\Lambda>\Lambda_{c}$, since only one bounded resonance remains, the oscillator may behave like a typical driven damped oscillator with one caveat. It still has a runaway solution, not seen in the Ohmic case.

\section{Higher-order supra-Ohmic systems}\label{S:kgbertdds}
It is interesting to briefly discuss the higher-order supra-Ohmic systems. Suppose the bath spectral density has the form
\begin{equation}
	J(\kappa)=\kappa^{4}\,I(\kappa)\,,
\end{equation}
i.e., $\lambda=5$, and then the equation corresponding to \eqref{E:krhsgfsd} has the form
\begin{align}\label{E:fkghbjdsd}
	m_{\textsc{p}}^{\vphantom{2}}\,\ddot{\chi}(t)&+m_{\textsc{p}}^{\vphantom{2}}\omega_{\textsc{p}}^{2}+e^{2}\Gamma^{(\textsc{e})}(0)\,\chi^{(4)}(t)-e^{2}\int_{0}^{t}\!ds\;\Gamma^{(\textsc{e})}(t-s)\,\chi^{(5)}(s)\\
	&\qquad-e^{2}\Bigl[\Gamma^{(\textsc{e})}(t)\,\chi^{(4)}(0)+\dot{\Gamma}^{(\textsc{e})}(t)\,\dddot{\chi}(0)+\ddot{\Gamma}^{(\textsc{e})}(t)\,\ddot{\chi}(0)+\dddot{\Gamma}^{(\textsc{e})}(t)\,\dot{\chi}(0)+\partial^{4}_{t}\Gamma^{(\textsc{e})}(t)\,\chi(0)\Bigr]=e\,\xi(t)\,.\notag
\end{align}
In addition to the supra-Ohmic features we have described hitherto, namely, 1) appearance of the higher-order time derivatives of $\chi(t)$ in the nonlocal expression, and 2) presence of additional initial conditions in the equation of motion, this equation of motion has an extra term like $e^{2}\Gamma^{(\textsc{e})}(0)\,\chi^{(4)}(t)$. Two other similar terms $e^{2}\ddot{\Gamma}^{(\textsc{e})}(0)\,\ddot{\chi}(t)$ and $e^{2}\partial_{t}^{4}\Gamma^{(\textsc{e})}(0)\,\chi(t)$ have been absorbed into the mass and the natural frequency renormalization respectively. Thus it seems that there is no corresponding parameter to absorb the $e^{2}\Gamma^{(\textsc{e})}(0)\,\chi^{(4)}(t)$ term. It is not clear whether this may or may not cause an issue. We probably do not need to worry the limit $\Gamma^{(\textsc{e})}(0)\propto\delta(0)$ because it belongs to the runaway regime.

Let us take a look at the Laplace transformation of \eqref{E:fkghbjdsd}, which is given by
\begin{equation}
	\tilde{\chi}(\sigma)=\tilde{d}_{1}(\sigma)\,\chi(0)+\tilde{d}_{2}(\sigma)\,\dot{\chi}(0)+\frac{e}{m_{\textsc{p}}}\,\tilde{d}_{3}(\sigma)\,\tilde{\xi}(\sigma)\,,
\end{equation}
where
\begin{align}
	\tilde{d}_{1}(\sigma)&=\frac{[1-8\pi\gamma\,\ddot{\Gamma}^{(\textsc{e})}(0)]\,\sigma}{\sigma^{2}+\omega_{\textsc{p}}^{2}-8\pi\gamma\,\sigma^{5}\tilde{\Gamma}^{(\textsc{e})}(\sigma)+8\pi\gamma\,\sigma^{4}\Gamma^{(\textsc{e})}(0)}\,,\\
	\tilde{d}_{2}(\sigma)&=\frac{1-8\pi\gamma\,\ddot{\Gamma}^{(\textsc{e})}(0)}{\sigma^{2}+\omega_{\textsc{p}}^{2}-8\pi\gamma\,\sigma^{5}\tilde{\Gamma}^{(\textsc{e})}(\sigma)+8\pi\gamma\,\sigma^{4}\Gamma^{(\textsc{e})}(0)}\,,\\
	\tilde{d}_{3}(\sigma)&=\frac{\tilde{d}_{2}(\sigma)}{1-8\pi\gamma\ddot{\Gamma}^{(\textsc{e})}(0)}\,,
\end{align}
where we have used the fact that $\dot{\Gamma}^{(\textsc{e})}(0)=0=\dddot{\Gamma}^{(\textsc{e})}(0)$. Surprisingly, the solution still depends on two initial conditions, not four, but the denominator of $\tilde{d}_{i}(\sigma)$ does contain the contribution from the previously mentioned $e^{2}\Gamma^{(\textsc{e})}(0)\,\chi^{(4)}(t)$ term. Again, should it be there? It turns out that its presence is nonetheless needed for stability. Following our previous discussions, we examine the large $\sigma$ limit of the denominator of $\tilde{d}_{i}(\sigma)$. In the limit $\sigma\to\infty$, the Laplace transform $\tilde{\Gamma}^{(\textsc{e})}(\sigma)$ is dual to the small time limit $\Gamma^{(\textsc{e})}(t)$, which is approximately given by
\begin{equation}
	\lim_{t\to0}\Gamma^{(\textsc{e})}(t)=\Gamma^{(\textsc{e})}(0)+\frac{1}{2}\,\ddot{\Gamma}^{(\textsc{e})}(0)\,t^{2}+\cdots\,.
\end{equation}
Then the Laplace transform is
\begin{equation}
	\lim_{\sigma\to\infty}\tilde{\Gamma}^{(\textsc{e})}(\sigma)=\frac{1}{\sigma}\,\Gamma^{(\textsc{e})}(0)+\frac{1}{\sigma^{3}}\,\ddot{\Gamma}^{(\textsc{e})}(0)+\cdots\,,
\end{equation}
so that we have
\begin{align}
	\sigma^{2}+\omega_{\textsc{p}}^{2}-8\pi\gamma\,\sigma^{5}\tilde{\Gamma}^{(\textsc{e})}(\sigma)+8\pi\gamma\,\sigma^{4}\Gamma^{(\textsc{e})}(0)&=\sigma^{2}+\omega_{\textsc{p}}^{2}-8\pi\gamma\,\sigma^{5}\Bigl[\frac{1}{\sigma}\,\Gamma^{(\textsc{e})}(0)+\frac{1}{\sigma^{3}}\,\ddot{\Gamma}^{(\textsc{e})}(0)+\cdots\Bigr]+8\pi\gamma\,\sigma^{4}\Gamma^{(\textsc{e})}(0)\notag\\
	&=\bigl[1-8\pi\gamma\,\ddot{\Gamma}^{(\textsc{e})}(0)\bigr]\,\sigma^{2}+\mathcal{O}(\sigma^{0})\,.\label{E:psdbwzw}
\end{align}
Thus when $1-8\pi\gamma\,\ddot{\Gamma}^{(\textsc{e})}(0)<0$, Eq.~\eqref{E:psdbwzw} grows unbounded to   minus infinity, so there is at least one positive real root. That is, $\tilde{d}_{i}(\sigma)$ will have at least one real pole such that the fundamental solutions $d_{1}(t)$ and $d_{2}$ are running away. If the contribution of the term $e^{2}\Gamma^{(\textsc{e})}(0)\,\chi^{(4)}(t)$ is not included in the denominator of $\tilde{d}_{i}(\sigma)$, then the stability condition will be inconsistent with the previous discussions. But it is not clear how to interpret  the $e^{2}\Gamma^{(\textsc{e})}(0)\,\chi^{(4)}(t)$ term in the equation of motion, \eqref{E:fkghbjdsd}.

Finally, we remark that when $\lambda$ is an even integer, we cannot reduce the nonlocal term in the Langevin equation into a local form, even in the formal Markovian limit $\Lambda\to\infty$. For example, we take a look at the Markovian case $J(\kappa)=\kappa^{2}$, and then
\begin{equation}
	G_{R,0}^{(\textsc{e})}(\tau)=i\,\theta(\tau)\int_{0}^{\infty}\!\frac{d\kappa}{2\pi}\;\frac{\kappa^{2}}{2\pi}\Bigl[e^{-i\kappa\tau}-e^{+i\kappa\tau}\Bigr]=-\theta(\tau)\,\frac{1}{\pi^{2}\tau^{3}}\,.
\end{equation}
It is a polynomial in $\tau$, so it will not give a delta function by way of integration by parts, rendering  the procedure of renormalization ambiguous. To understand better why it has this form, let us put back the spatial dependence in the retarded Green's function in the formal Markovian limit
\begin{equation}
	G_{R,0}^{(\textsc{e})}(\tau,\bm{R})=i\,\theta(\tau)\int_{0}^{\infty}\!d\kappa\;\frac{\kappa^{2}}{2}\!\int\!\frac{d\Omega}{(2\pi)^{3}}\;\Bigl[e^{i\bm{k}\cdot\bm{R}-i\kappa\tau}-e^{-i\bm{k}\cdot\bm{R}+i\kappa\tau}\Bigr]=-\frac{\theta(\tau)}{\pi^{2}}\frac{\tau}{(\tau^{2}-R^{2})^{2}}\,,
\end{equation}
with $R=\lvert\bm{R}\rvert$. It has support in the whole spacetime, not just limited to the lightcone, so it has contributions outside the lightlike interval. This can be a bad sign because the corresponding Hadamard function at the coincident spatial limit may take the form of a delta function. The Hadamard function $G_{H,0}^{(\textsc{e})}(\tau,\bm{R})$ is given by
\begin{align}
	G_{H,0}^{(\textsc{e})}(\tau,\bm{R})&=-\frac{i}{2R}\int_{0}^{\infty}\!\frac{d\kappa}{2\pi}\;\frac{\kappa}{4\pi}\coth\frac{\beta\kappa}{2}\Bigl[e^{-i\kappa(\tau-R)}-e^{+i\kappa(\tau-R)}-e^{-i\kappa(\tau+R)}+e^{+i\kappa(\tau+R)}\Bigr]\label{E:uyfdvgjdf}\\
	&=\frac{i}{8\pi^{2}\beta^{2}R}\Bigl[-\psi^{(1)}(+i\,\frac{\tau-R}{\beta})+\psi^{(1)}(-i\,\frac{\tau-R}{\beta})-\psi^{(1)}(-i\,\frac{\tau+R}{\beta})+\psi^{(1)}(+i\,\frac{\tau+R}{\beta})\Bigr]\,,\notag
\end{align}
and in the limit $R\to0$, it reduces to
\begin{align}
	G_{H,0}^{(\textsc{e})}(\tau,\bm{0})&=-\frac{1}{4\pi^{2}\beta^{3}}\Bigl[\psi^{(2)}(+i\,\frac{\tau}{\beta})+\psi^{(2)}(-i\,\frac{\tau}{\beta})\Bigr]\,.
\end{align}
This may not look like a delta function at all. However, in the zero temperature limit $\beta\to\infty$, we have
\begin{align}
	G_{H,0}^{(\textsc{e})}(\tau,\bm{R})&=-\frac{i}{2R}\int_{0}^{\infty}\!\frac{d\kappa}{2\pi}\;\frac{\kappa}{4\pi}\Bigl[e^{-i\kappa(\tau-R)}-e^{+i\kappa(\tau-R)}-e^{-i\kappa(\tau+R)}+e^{+i\kappa(\tau+R)}\Bigr]\notag\\
	&=\frac{1}{8\pi R}\frac{\partial}{\partial\tau}\Bigl[\delta(\tau-R)-\delta(\tau+R)\Bigr]\,,\label{E:fkghveweser}
\end{align}
and now we see the delta function emerges. In particular, in the coincident spatial limit, it becomes
\begin{equation}
	G_{H,0}^{(\textsc{e})}(\tau,\bm{0})=-\frac{1}{4\pi}\,\delta''(\tau)\,,\label{E:zzfkghsgf}
\end{equation}
a highly singular entity. It is very non-intuitive. From these results we notice that the functional forms of the dissipation and the noise kernels swap,  posing  interpretation difficulty when $\lambda$ is an even integer.

\end{document}